\documentstyle [a4,11pt]{article}
\def\sign(#1){(\!-\!1)^{#1}}
\def\binom(#1,#2){ (\!\!
	 \begin{array}{c} #1 \\ #2 \end{array}\!\! ) }
\def\plus{\!+\!}
\def\minus{\!-\!}
\def\mydot{\!\!\cdot\!}
\def\nn{\nonumber \\ &&}
\def\nne{\nonumber \\ & = &}

\def\Li{{\rm Li}}
\def\scrarrow#1{{\overrightarrow{\scriptstyle \! #1}}}

\begin{document}
\begin{titlepage}
\hfill FTUAM 98-7

\hfill NIKHEF-98-14

\begin{center}
{\huge Harmonic sums, Mellin transforms and Integrals} \\ [8mm]
{\large J. A. M. Vermaseren} \\ [4mm]
Departamento de F\'\i sica Te\'orica, C-XI, \\
Universidad Aut\'onoma de Madrid, \\
Canto Blanco,\\
E-28034 Madrid, SPAIN. \\ [3mm]
and \\ [3mm]
NIKHEF \\
P.O.Box 41882 \\
NL-1009DB, Amsterdam \\ [10mm]
\today \\ [10mm]
\end{center}
 
\begin{abstract}\noindent
This paper describes algorithms to deal with nested symbolic sums over 
combinations of harmonic series, binomial coefficients and denominators. In 
addition it treats Mellin transforms and the inverse Mellin transformation 
for functions that are encountered in Feynman diagram calculations. 
Together with results for the values of the higher harmonic series at 
infinity the presented algorithms can be used for the symbolic evaluation 
of whole classes of integrals that were thus far intractable. Also many of 
the sums that had to be evaluated seem to involve new results. Most of the 
algorithms have been programmed in the language of FORM. The resulting set 
of procedures is called SUMMER.

\end{abstract}
\end{titlepage}

\section{Introduction}

The computation of Feynman diagrams has confronted physicists with classes 
of integrals that are usually hard to be evaluated, both analytically and 
numerically. Also the newer techniques applied in the more popular computer 
algebra packages do not offer much relief. Therefore it is good to 
occasionally study some alternative methods to come to a result. In the 
case of the computation of structure functions in deep inelastic scattering 
one is often interested in their Mellin moments. Each individual moment can 
be computed directly in ways that are much easier than computing the whole 
structure function and taking its moments afterwards. There exist however 
also instances in the literature in which all moments were evaluated in a 
symbolic way~\cite{First}~\cite{Gross}~\cite{Madrid}~\cite{KandK}. Once all 
positive even moments are known, one can reconstruct the complete structure 
functions. Hence such calculations contain the full information and are in 
principle as valuable as the direct evaluation of the complete integrals. 
In these calculations the integrals become much simpler at the cost of 
having to do a number of symbolic sums over harmonic series. The draw-back 
of the method is that although much effort has been put in improving 
techniques of integration over the past years, very little is known about 
these classes of sums. A short introduction is given for instance in 
ref~\cite{Knuth}. In addition such calculations are of a nature that one 
needs to do them usually by means of a computer algebra program. This means 
that when algorithms are developed, they should be suitable for 
implementation in the language of such programs.

This paper describes a framework in which such calculations can be done. As 
such it gives a consistent notation that is suited for a computer program. 
It shows a number of sums that can be handled to any level of complexity 
and describes an implementation of them in the language of the program 
FORM~\cite{form}. Then the formalism is applied to the problem of Mellin 
transforms of a class of functions that traditionally occurs in the 
calculation of Feynman diagrams. This in its turn needs harmonic series in 
infinity and hence there is a section on this special case. Next the 
problem of the inverse Mellin transform is dealt with. With the results of 
the series at infinity one can suddenly evaluate a whole class of integrals 
symbolicaly. This is explained in the next section where some 
examples are given.

The paper is finished with a number of appendices. They describe the 
details of some of the algorithms and their implementation. Additionally 
there is an appendix with lists 
of symbolic sums that are not directly treated by the `general' 
algorithms. These sums were obtained during various phases of the project 
and many of them do not seem to occur in the literature.

\section{Notations}

The notation that is used for the various functions and series in this 
paper is closely related to how useful it can be for a computer 
program. This notation stays as closely as possible to existing 
ones. The harmonic series is defined by
\begin{eqnarray}
    S_m(n)    & = & \sum_{i=1}^n \frac{1}{i^m} \\
    S_{-m}(n) & = & \sum_{i=1}^n \frac{\sign(i)}{i^m}
\end{eqnarray}
in which $m > 0$. One can define higher harmonic series by
\begin{eqnarray}
    S_{m,j_1,\cdots,j_p}(n) & = & \sum_{i=1}^n \frac{1}{i^m}
            S_{j_1,\cdots,j_p}(i) \\
    S_{-m,j_1,\cdots,j_p}(n) & = & \sum_{i=1}^n \frac{\sign(i)}{i^m}
            S_{j_1,\cdots,j_p}(i)
\end{eqnarray}
with the same conditions on $m$. The $m$ and the $j_i$ are referred to as 
the indices of the harmonic series.
Hence
\begin{equation}
    S_{1,-5,3}(n) = \sum_{i=1}^n\frac{1}{i}
                    \sum_{j=1}^i\frac{\sign(j)}{j^5}
                    \sum_{k=1}^j\frac{1}{k^3}
\end{equation}
In the literature the alternating sums are usually indicated by a bar over 
the index. The advantage of this notation is that it can be extended easily 
for use in a computer algebra program, eg.:

\begin{center}
$S_{i_1,\cdots,i_m}(n) \rightarrow$ {\tt S(R(i1,...,im),n)}.
\end{center}

\noindent Such objects can be easily manipulated in the more modern versions of 
the program FORM.

The argument of a harmonic series which has only positive 
indices can be doubled with the formula:
\begin{equation}
\label{eq:doubling}
	S_{m_1,\cdots,m_p}(n) = \sum_{\pm} 2^{m_1+\cdots+m_p-p}
			 S_{\pm j_1,\cdots,\pm j_p}(2n)
\end{equation}
in which the sum is over all $2^p$ combinations of $+$ and $-$ signs.

The weight of a harmonic series is defined as the sum of the absolute values 
of its indices.
\begin{equation}
    W(S_{j_1,\cdots,j_m}(n)) = \sum_{i=1}^m |j_i|
\end{equation}

For any positive weight $w$ there are $2\times 3^{w\minus 
1}$ linearly independent harmonic series. 
The fact that for each next weight there are three times as many can be seen 
easily: One can extend the series of the previous weight either by putting 
an extra index $1$ or $-1$ in front, or by raising the absolute value of 
the first index by one.
 
The set of all harmonic series with the same weight is called the 
`natural' basis for that weight. 

The extended weight of the compound object of a series and denominators is 
the weight of the series plus the number of powers of denominators that are 
identical to the argument of the series. Hence $S_{1,-5,3}(n)/n^4$ has the 
extended weight 13.

The total weight of a term is the sum of all extended weights of all the 
series in that term. Hence the total weight of $S_{2,3}(n)S_{-2}(m)$ is 7.

The value 0 for an index is reserved for an application that is typical for 
computers. If the results of a given weight need to be tabulated, the above 
notation would require a table in which the number of indices is not fixed. 
This can be remedied by a modified notation which is 
only used in specific stages of the program. An index that is zero which is 
followed by an index that is nonzero indicates that one should be added to 
the absolute value of the nonzero index. Hence:
\begin{eqnarray}
\label{eq:zeroes}
	S_{0,1,0,0,-1,1}(n) & = & S_{2,-3,1}(n)
\end{eqnarray}
This way one can express all series of weight $w$ into functions with $w$ 
indices of which the first $w\minus 1$ can take the values $1$, $0$ and 
$-1$, while the last one can take the values $1$ and $-1$.

Consider the following identity which can be obtained by exchanging the 
order of summation:
\begin{equation}
    S_{j,k}(n)+S_{k,j}(n) = S_j(n)S_k(n)+S_{j \& k}(n)
\end{equation}
in which the pseudo addition operator $\&$ adds the absolute values and 
gives the result a positive value if $j$ and $k$ have the same sign and 
otherwise the result will have a negative value. One could select a basis 
in which one keeps products of harmonic series with as 
simple a weight as possible. The above equation would indicate that in that 
case one of the left terms should be excluded from the basis in favor of 
the first term on the right hand side. Although the choice of which of the 
higher harmonic series to keep and which to drop in favor of the product 
terms is not unique, there are cases in which such a basis is to be 
preferred. In particular when $n\rightarrow\infty$ one can choose a basis 
in which all divergent objects are expressed as powers of $S_1(\infty)$ 
multiplied by finite harmonic series. In general however the summation 
formulae are much simpler in the natural basis in which each element is a 
single higher harmonic series. This can be seen rather easily when looking at 
the sum:
\begin{eqnarray}
 \sum_{i=1}^n \frac{S_j(i)S_k(i)}{i^m} & = &
    \sum_{j=1}^n\frac{S_{j,k}(i)+S_{k,j}(i)-S_{j \& k}(i)}{i^m}
        \nonumber \\ & = &
    S_{m,j,k}(n)+S_{m,k,j}(n)-S_{m,j\&k}(n)
\end{eqnarray}
In order to be able to do the sum one has to convert to the natural basis 
anyway. After the summation one would have to convert back to whatever 
other basis one happens to work with. Appendix A presents an 
algorithm by which combinations of harmonic series with the same argument 
can be expressed in the natural basis.

Additionally there can be denominators containing the summation parameter. 
There arises immediately a problem when 
there is more than one denominator. Traditionally one can split 
the fractions with
\begin{equation}
    \frac{1}{i+a}\frac{1}{i+b} = \frac{1}{b-a}(\frac{1}{i+a}-\frac{1}{i+b})
\end{equation}
Unfortunately this formula is not correct when $a=b$. Because often there 
will be nested sums and sums with symbolic parameters, $a$ and $b$ can be 
functions of summation or other parameters and hence it will not be obvious 
when $a=b$. In FORM this can be repaired in principle with one of the special 
functions:
\begin{eqnarray}
\label{eq:onemindelta}
	\frac{1}{i\plus a}\frac{1}{i\plus b} & = &
				\delta_{a,b}\frac{1}{(i\plus a)^2}
				+(1-\delta_{a,b})\frac{1}{b-a}(\frac{1}{i+a}-\frac{1}{i+b})
\end{eqnarray}
Here $\delta_{a,b}=1$ when $a=b$ and zero otherwise. In the language of 
FORM it is represented by the built in object \verb:delta_:. 
Unfortunately this form of the partial fractioning is not very 
useful, because it still evaluates into terms involving {1/(b-a)} in 
which $a$ can be equal to $b$. Hence an even more 
complicated form is needed:
\begin{eqnarray}
\label{eq:assumeint}
	\frac{1}{i\plus a}\frac{1}{i\plus b} & = &
				\delta_{a,b}\frac{1}{(i\plus a)^2}
				+(\theta(a\minus b\minus 1)+\theta(b\minus a\minus 1))
					\frac{1}{b-a}(\frac{1}{i+a}-\frac{1}{i+b})
\end{eqnarray}
in which one has to assume that $a$ and $b$ only take integer values. The 
function $\theta(x)$ (in FORM \verb:theta\_(x):) is zero when $x$ is 
negative and one when $x$ is zero or positive. These $\theta$-functions 
fulfill the r\^ole of conditions like $a\ge b\plus 1$ plus $b\ge a\plus 1$ 
and are worked out first. Hence this should not be read as $0/0$ for the 
case that $a=b$. The complete and proper equation would involve a new 
function:
\begin{eqnarray}
	\frac{1}{i\plus a}\frac{1}{i\plus b} & = &
				\delta_{a,b}\frac{1}{(i\plus a)^2}
				+(\theta'(a\minus b)+\theta'(b\minus a))
					\frac{1}{b-a}(\frac{1}{i+a}-\frac{1}{i+b})
\end{eqnarray}
with $\theta'(x)$ (in FORM \verb:thetap\_(x):) is one when $x > 0$ and zero 
when $x \le 0$. Actually $\theta'(x) = 1\minus\theta(-x)$, but 
this cannot be used for the same reason that the 
equation~(\ref{eq:onemindelta}) 
could not be used. 
Because it is rather complicated to manipulate both the functions 
$\theta(x)$ and $\theta'(x)$ simultaneously, 
and because one has almost always integer values 
of the parameter, the computer program uses mostly the 
formula~(\ref{eq:assumeint}) which 
assumes the integer values. It should be clear that much attention should 
be given to theta and delta functions, their combinations and their 
interactions with summations.

\section{Synchronization}

When one has to do sums over a combination of objects one of the problems 
is that such objects do not always have identical arguments. If this is the 
case one would have to program many more sums than often is necessary. 
Whenever it is possible one should `synchronize' the arguments. This means 
that one tries to make the arguments of the various harmonic series, the 
denominators and the factorials equal to each other. This 
can be illustrated with one harmonic series and one denominator:
\begin{eqnarray}
    \sum_{i=1}^n \frac{S_1(i\plus 1)}{i} & = &
        \sum_{i=1}^n \frac{S_1(i)}{i} + \sum_{i=1}^n\frac{1}{i\plus 1}
            \frac{1}{i}  \nonumber \\ & = &
        \sum_{i=1}^n \frac{S_1(i)}{i} + \sum_{i=1}^n\frac{1}{i}
            -\sum_{i=1}^n\frac{1}{i\plus 1}
\end{eqnarray}
In this equation and the sequel it is assumed that the left most index is 
positive. If it is negative there will be the extra $\sign(i)$ and one 
has to be more careful with the signs of the terms, but the principle is 
always the same.

Of course, when the difference between some arguments is symbolic like in 
$S_1(i\plus k)/i$, such tricks do not work, but for differences that are 
integer constants one can define a scheme that converges. Let $m$ be 
a positive integer constant in the remaining part of this section. In that 
case one can write:
\begin{eqnarray}
    \frac{S_{j,r_1,\cdots,r_s}(i\plus m)}{i} & = &
    \frac{S_{j,r_1,\cdots,r_s}(i\plus m\minus 1)}{i}
    + \frac{S_{r_1,\cdots,r_s}(i\plus m)}{i(i\plus m)^j}
\end{eqnarray}
The partial fractioning of the denominators in the last term results in 
terms that have only a power of $1/(i\plus m)$ and one term which has a 
factor $1/i$. This last term however has a simpler harmonic series in the 
numerator. Hence this relation defines a recursion that terminates. 
Similarly one can write:
\begin{eqnarray}
     \frac{S_{j,r_1,\cdots,r_s}(i)}{i\plus m} & = &
     \frac{S_{j,r_1,\cdots,r_s}(i\plus 1)}{i\plus m}
    -\frac{S_{r_1,\cdots,r_s}(i\plus 1)}{(i\plus m)(i\plus 1)^j}
\end{eqnarray}
and partial fractioning results again in terms in which the arguments 
either are the same, or closer to each other, or the harmonic series has 
become simpler. 

Next is the interaction between two harmonic series:
\begin{eqnarray}
    S_{j,r_1,\cdots,r_s}(i) S_{p_1,\cdots,p_q}(i\plus m)
        & = &
    S_{j,r_1,\cdots,r_s}(i\plus 1) S_{p_1,\cdots,p_q}(i\plus m) \nn
    -S_{r_1,\cdots,r_s}(i\plus 1) S_{p_1,\cdots,p_q}(i\plus m)
        \frac{1}{(i\plus 1)^j}
\end{eqnarray}
This relation defines, in combination with the previous two equations, also 
a proper recursion. In the last term one can synchronize the argument of 
the second harmonic series with that of the denominator, giving 
(potentially many) terms with either $1/(i\plus m)$ or an argument that is 
closer to $i\plus 1$. In all cases the arguments are at least one closer to 
each other. In addition some of the harmonic series have become simpler.

Once two harmonic series have the same arguments this product can be 
rewritten into the basis of single higher harmonic series (see appendix A). 
Hence products of more than two harmonic series with different arguments 
can be dealt with successively.

At this point there can still be factorials. The beginning is easy:
\begin{eqnarray}
    \frac{1}{i\ (i\plus m)!} & = &
             \frac{1}{i\ (i\plus m)\ (i\plus m\minus 1)!} \nonumber \\ & = 
    & \frac{1}{m}\frac{1}{i\ (i\plus m\minus 1)!}
     -\frac{1}{m}\frac{1}{(i\plus m)!}
\end{eqnarray}
and
\begin{eqnarray}
    \frac{1}{(i\plus m)\ i!} & = &
             \frac{i\plus 1}{(i\plus m)\ (i\plus 1)!} \nonumber \\ & = 
    & \frac{1}{(i\plus 1)!}-\frac{m\minus 1}{(i\plus m)\ (i\plus 1)!}
\end{eqnarray}
Because of these two equations one can also synchronize combinations of 
harmonic series and factorials.

The problem is that usually one cannot do very much with the product of two 
factorials. This means that if one has more than one factorial, one may be 
left with factorials with different arguments.

Another problem exists with arguments of the type $i$ versus arguments of 
the type $n\minus i$. These can of course not be synchronized completely, 
but if $n$ is the upper limit of the summation over $i$, one can try to make 
a synchronization that excludes other nonsymbolic constants. This is 
slightly more complicated than what was done before:
\begin{eqnarray}
    \frac{1}{n\minus i}S_{j,r_1,\cdots,r_s}(i\plus 1) &=&
        \frac{1}{n\minus i}S_{j,r_1,\cdots,r_s}(i)
        +\frac{1}{(n\minus i)(i\plus 1)^j}S_{r_1,\cdots,r_s}(i\plus 1)
\end{eqnarray}
Partial fractioning of the last term will leave something simpler.
Similarly there is:
\begin{eqnarray}
    \frac{1}{n\minus i}S_{j,r_1,\cdots,r_s}(i\minus 1) &=&
        \frac{1}{n\minus i}S_{j,r_1,\cdots,r_s}(i)
        -\frac{1}{(n\minus i)(i\minus 1)^j}S_{r_1,\cdots,r_s}(i\minus 1)
\end{eqnarray}
For two $S$-functions one can write:
\begin{eqnarray}
    S_{j,r_1,\cdots,r_s}(n\minus i)S_{k,p_1,\cdots,p_s}(i\plus 1) &=&
    S_{j,r_1,\cdots,r_s}(n\minus i)S_{k,p_1,\cdots,p_s}(i) \nn
    +\frac{1}{(i\plus 1)^k}S_{j,r_1,\cdots,r_s}(n\minus i)
                S_{p_1,\cdots,p_s}(i\plus 1)
        \nonumber \\ &=&
    S_{j,r_1,\cdots,r_s}(n\minus i)S_{k,p_1,\cdots,p_s}(i) \nn
    +\frac{1}{(i\plus 1)^k}S_{j,r_1,\cdots,r_s}(n\minus i\minus 1)
                S_{p_1,\cdots,p_s}(i\plus 1) \nn
    +\frac{1}{(i\plus 1)^k(n\minus i)^j}
                S_{r_1,\cdots,r_s}(n\minus i)
                S_{p_1,\cdots,p_s}(i\plus 1)
\end{eqnarray}
Again partial fractioning of the last term leads to a simpler object.
One can derive equivalent relations for combinations involving factorials. 
In this case also pairs of factorials can be dealt with:
\begin{eqnarray}
    \frac{1}{(n\minus i)!\ (i\plus 1)!} & = &
        \frac{1}{(n\minus i\minus 1)!\ i!}
        \frac{1}{n\plus 1}(\frac{1}{n\minus i}+\frac{1}{i\plus 1})
        \nonumber \\ &=&
        \frac{1}{n\plus 1}(\frac{1}{(n\minus i)!\ i!}
            +\frac{1}{(n\minus i\minus 1)!\ (i\plus 1)!} )
\end{eqnarray}
All the above relations can be combined into one recursion that 
leaves all $S$-functions, all denominators and at least one factorial
properly synchronized. Additionally one has a proper adjustment to the 
boundaries of the summation, and therefore the factorials can often be 
combined into binomial coefficients.


\section{Mellin Transforms}

\label{sec:mellin}
The Mellin operator $M$ is defined by
\begin{equation}
\label{eq:mellin}
	M(f(x)) = \int_0^1dx\ x^m f(x)
\end{equation}
and the operator $M^+$ by
\begin{equation}
	M^+(f(x)) = \int_0^1dx\ x^m \frac{f(x)}{(1\minus x)_+}
\end{equation}
with
\begin{equation}
	\int_0^1 dx\frac{f(x)}{(1\minus x)_+} =
		\int_0^1 dx\frac{f(x)-f(1)}{1\minus x}
\end{equation}
when $f(1)$ is finite.
When there is a power of $\ln(1\minus x)$ present it becomes
\begin{eqnarray}
	M^+((\ln(1\minus x))^kf(x)) & = &
			 \int_0^1dx\ x^m \frac{(\ln(1\minus x))^kf(x)}{(1\minus x)_+} \\
	& = & \int_0^1 dx ( f(x)-f(1) )\frac{(\ln(1\minus x))^k}{1\minus x}
\end{eqnarray}
These are the traditional operations. In the literature one often defines 
the transform shifted over one as in
\begin{equation}
	M(f(x)) = \int_0^1dx\ x^{N-1} f(x)
\end{equation}
In the context of this paper the notation will be the one of 
equation~(\ref{eq:mellin}). 
The `Mellin parameter' is given in that case in lower 
case variables. Hence the translation to the shifted notation 
should be of the nature $n\rightarrow N\minus 1$.

For Mellin transforms of formulas resulting from Feynman diagrams one has 
to consider the transforms of functions that are 
combinations of $1/(1\minus x)_+$, $1/(1+x)$, $\ln(x)$, $\ln(1\plus x)$,
$\ln(1\minus x)$, powers of these logarithms, and various polylogarithms of 
which the arguments are rational functions of $x$. 
Powers of $x$ just change the moment of the function. Hence they do not have 
to be considered. Additionally one can always assume that either 
$1/(1\minus x)_+$ or $1/(1+x)$ is present, because the functions without 
such a term can be written as two functions in the class that is being 
considered:
\begin{eqnarray}
	1 & = & \frac{1}{1\plus x} + \frac{x}{1\plus x}
\end{eqnarray}

The algorithm that obtains the Mellin transform of any 
combination of such functions is rather direct. Consider the following 
steps:
\begin{enumerate}
\item If there is a power of $1/(1\minus x)$ or $1/(1\plus x)$, replace 
it by a sum according to the formulas
\begin{eqnarray}
	\frac{x^m}{1\minus x} & = & \sum_{i=m}^{\infty}x^i \\
	\frac{x^m}{1\plus x} & = & \sign(m)\sum_{i=m}^{\infty}\sign(i)x^i
\end{eqnarray}
\item If the function to be transformed contains powers of $\ln(1\minus 
x)$, split it into its powers of $\ln(1\minus x)$ and $F(x)$ 
which represents the rest and has a finite value at $x=1$. Then one writes
\begin{eqnarray}
	\int_0^1dx\ x^m\ln^p(1\minus x)\ F(x) & = &
	\int_0^1dx\ x^m\ln^p(1\minus x)\ (F(x)-F(1)) \nonumber \\ &&
	+F(1)\int_0^1dx\ x^m\ln^p(1\minus x)
\end{eqnarray}
\item The Mellin transform of just a power of $\ln(1\minus x)$ 
can be replaced immediately using the formula
\begin{eqnarray}
	\int_0^1dx\ x^m\ln^p(1\minus x) & = & \frac{\sign(p)p!}{m\plus 1}
			S_{1,\cdots,1}(m\plus 1)
\end{eqnarray}
in which the $S$-function has $p$ indices that are all 1. This avoids 
divergence problems during the next step. Similarly one can apply:
\begin{equation}
	\int_0^1 dx\ x^m \ln^p(x) = \frac{\sign(p)p!}{(m\plus 1)^{p\plus 1}}
\end{equation}
when there is only a power of $\ln(x)$ left, but this step is not 
essential; it only makes the algorithm a bit faster. Due to the powers of 
$x$ there will be no divergence problems near $x=0$.
\item Do a partial integration on the powers of $x$. 
Because of the second step, the values at 
$x=0$ and $x=1$ never present any problems.
\item If there is only a power of $x$ left one can integrate and the 
integration phase is finished. Otherwise one should repeat the previous 
steps until all functions have been broken down. Note that for this to work 
all functions have to break down properly. Hence one cannot use fractional 
powers of the functions involved.
\item At this point the terms may contain nested sums, either to a finite 
upper limit or to infinity. These sums do not present any complications 
once products of two $S$-functions with identical arguments 
can be combined into elements of the natural basis (see appendix A).
\end{enumerate}
The main complication in the above algorithm is the treatment of the 
infinities that may arise in the summations. Many of the terms develop a 
divergence. These are all of a rather soft nature and hence their 
regularization is relatively easy. All divergences in the sums are of a 
logarithmic nature and hence, if one considers the sum to go to a rather 
large integer $L$, the divergent sums behave like powers of $\ln L$ up to 
terms of order $1/L$\footnote{In principle there is also an Euler constant, 
but when the logarithms cancel, also the Euler constants cancel and hence 
they are not considered here}. Because all transforms should be finite the 
terms in $\ln L$ should cancel. After that one can safely take the limit 
$L\rightarrow \infty$. Taking this all in consideration, all sums that 
contain a divergence can be rewritten into powers of one single basic 
divergent sum ($S_1(\infty)$) and finite terms. After that there are no 
more problems of this nature.

The result of the above algorithm is an expression with many harmonic 
series of which the argument is a function of $m$ and others of which the 
argument is infinity. These last sums are treated in the next section.


\section{Values at infinity}

In the previous section the results of the Mellin transforms were harmonic 
series in the Mellin parameter $m$ and harmonic series at infinity. In 
order to solve the problem completely one has to find the values for these 
series at infinity. After all they represent finite numbers and the number 
of series is much larger than the number of transcendental numbers that 
occur once they are evaluated. The sums to be considered are related to 
the Euler-Zagier sums~\cite{Euler}~\cite{Zagier} which are defined as
\begin{equation}
\zeta(s_1,\ldots,s_k;\sigma_1,\ldots,\sigma_k)=\sum_{n_j>n_{j+1}>0}
\quad\prod_{j=1}^{k}\frac{\sigma_j^{n_j}}{n_j^{s_j}}\,.
\end{equation}
These sums are however not identical to the $S$-functions at infinity 
because for them the sum is
\begin{equation}
S_{s_1,\ldots,s_k}(\infty)=\sum_{n_j\ge n_{j+1} \ge 1}
\quad\prod_{j=1}^{k}\frac{[\sign(n_j)]_{s_j<0}}{n_j^{|s_j|}}\,.
\end{equation}
The notation $[\ \ ]_{s_1<0}$ indicates that this part is present only 
when $s_1 < 0$.
Here a method is presented to evaluate these sums that is completely 
different from the one in reference~\cite{Broadhurst}.

The first step in the evaluation of the sums is to express the sums as much 
as possible in terms of products of harmonic series with a lower weight. 
This can be done up to a point. One will always need a number of series 
with the weight one is considering. This step is basically the inverse of 
the algorithm of appendix A. It is harder to be implemented in a deterministic 
way, because the choice of the basis is not unique. But this can be solved 
in a different way as will be seen below.

Next there are two types of extra identities one can consider. The first 
set comes from looking at the series with only positive indices and 
applying the doubling formula~(\ref{eq:doubling}) to it. For all the series 
that are finite it makes no difference whether the argument is infinity or 
two times infinity. If the selected basis is such 
that all divergences are powers of $S_1(\infty)$ one only has to make the 
extra adjustment $S_1(2\infty) \rightarrow S_1(\infty)+\ln(2)$. This gives 
a number of extra equations that correspond to new relations between 
the series. Unfortunately this does not give enough relations, but some are 
interesting in their own right. For instance
\begin{eqnarray}
	S_m(\infty) & = & 2^{m-1}(S_m(2\infty)+S_{-m}(2\infty))
\end{eqnarray}
gives immediately the well known relations
\begin{eqnarray}
	S_{-m}(\infty) & = & -(1-2^{1-m})S_m(\infty)\ \ \ \ \ \ \ \ \ \ m > 1
			\\
	S_{-1}(\infty) & = & -\ln(2)
\end{eqnarray}

The more powerful consideration however is the following: Suppose one is 
summing over a square grid of size $n\times n$. Under what conditions is 
the sum over the upper right diagonal half of the square ($i_1+i_2>n$) zero 
in the limit $n\rightarrow\infty$? If this sum is zero, the product over 
two individual sums can be replaced by a sum over the lower left diagonal 
triangle ($i_1+i_2 \le n$). This leads to the following theorem:

\noindent {\sl Theorem:}
When not both $m_1 = 1$ and $k_1 = 1$ the following identity holds:
\begin{eqnarray}
\label{eq:infinities}
S_{m_1,\cdots,m_p}(\infty)S_{k_1,\cdots,k_q}(\infty) & = &
\lim_{n\to\infty}\sum_{i=1}^{n}
S_{m_1,\cdots,m_p}(n\minus i)S_{k_2,\cdots,k_q}(i)\frac{[\sign(i)]_{k_1<0}
}{i^{|k_1|}}
\end{eqnarray}
The proof is rather trivial, considering that all $m_i$ and 
$k_i$ are integers and that alternate series with $\sign(i)$ actually 
converge one power of $i$ better than they seem to at first sight. This can 
be seen when the terms are grouped in pairs. The sums can be estimated by 
integrals and the numerators can only give powers of logarithms. Hence the 
presence of at least three powers of denominators (excluding $m_1 = k_1 = 
1$) will make the limit go to zero.

The sum can be readily worked out with the algorithm described in appendix 
C.

Assume that all sums up to weight $n$ have been determined. 
The complete algorithm for weight $n+1$ is now:
\begin{enumerate}
\item Construct all pairs of $S$-functions for which the sum of the two 
weights is $n+1$.
\item Each pair is used to construct two equations (unless both 
$S$-functions have their first index equal to one in which case the second 
equation that would have been based on the above theorem is not made).
The first equation is made by taking $S^{(1)}S^{(2)}-S^{(1)}S^{(2)}$ and 
applying the routine (see appendix A) that converts the $S$-functions to 
the basis to the first pair. 
These are the `shuffle algebra' relations.
The second set of equations is created by taking 
$S^{(1)}S^{(2)}-S^{(1)}S^{(2)}$ and then applying the formula of the theorem 
to the first pair. After this the routine of appendix C is applied.
\item Substitute the values for the lower $S$-functions.
\item Eliminate now the `unknown' $S$-functions of weight $n+1$ as much as 
possible as if one is solving a linear set of equations (which is what it 
is). Apply the same set of substitutions that will eliminate the 
equations to the series that need to be evaluated.
\item Inspect the result and see which sums should be considered as new 
independent variables because they were not eliminated. If one insists on a 
given sum to be among the variable(s) not to be eliminated one can 
substitute it by a different variable before the elimination procedure.
\end{enumerate}
It is not so difficult to construct a program in the language of FORM that 
can execute this procedure all the way to $S$-functions of weight 7. Such a 
program takes just a few hours ($< 6$ without special optimizations) on a 
Pentium-II-300 processor. When a series diverges one uses the basic 
divergence $S_1(\infty)$ as if it were a regular variable. This presents no 
problems.

The variables that one needs at the different weights are:
$S_1(\infty)$, $\ln(2)$, $\zeta_2$, $\zeta_3$, $\Li_4(\frac{1}{2})$,
$\zeta_5$, $\Li_5(\frac{1}{2})$, $\Li_6(\frac{1}{2})$, $S_{-5,-1}(\infty)$, 
$\zeta_7$, $\Li_7(\frac{1}{2})$, $S_{-5,1,1}(\infty)$, $S_{5,-1,
-1}(\infty)$. The choice of the $S$-functions that remain is not unique. 
Here the selection is such that they contain as few indices as possible and 
are as convergent as possible. Numerical values for these quantities can be 
obtained by standard techniques.
\begin{eqnarray}
	\Li_4(1/2)                      & = & 0.51747906167389938633 \nonumber \\
	\Li_5(1/2)                      & = & 0.50840057924226870746 \nonumber \\
	\Li_6(1/2)                      & = & 0.50409539780398855069 \nonumber \\
	\Li_7(1/2)                      & = & 0.50201456332470849457 \nonumber \\
	S_{\minus 5,\minus 1}(\infty)   & = & 0.98744142640329971377 \nonumber \\
	-S_{\minus 5,1,1}(\infty)       & = & 0.95296007575629860341 \nonumber \\
	S_{5,\minus 1,\minus 1}(\infty) & = & 1.02912126296432453422
\end{eqnarray}
It should be noted however that according 
to the work by Broadhurst and Kreimer~\cite{Kreimer} most of these 
constants should not appear in the computation of massless Feynman 
diagrams. The first non-zeta constant should be $S_{5,3}(\infty)$ which is 
an object of weight 8. This indicates that in $x$-space the functions can 
only occur in such combinations that these constants cancel in Mellin 
space. Hence one may not need to know their values for many applications. 
In the case of massive Feynman diagrams the situation is different. The 
constant $\Li_4(1/2)$ does occur in the three loop corrections to the 
$g\minus 2$ of the electron~\cite{Remiddi}.

The results of the runs up to weight 7 have been tabulated and put in the 
FORM program. The main problem in making the tables is that the objects 
with identical weights may have different numbers of indices. Hence the 
notation of indices that are either $-1$, $1$ or $0$ of equation 
(\ref{eq:zeroes}) is used for the 
tables. The conversion to and from this notation is rather simple.


\section{Inverse Mellin Transforms}

If one can obtain a result in Mellin space (as a function of $n$) in 
principle it is possible to convert to the function in $x$-space. This is 
however a rather complicated operation. There exists some literature about 
it~\cite{Gross}~\cite{invmel} but it remains rather difficult. Also 
considering it as some type of Laplace transform does not give much 
relief~\cite{Laplace}. In many cases one can employ a different strategy. 
Given a result in Mellin space with a set of series, one can try to find a 
set of functions in $x$-space for which the Mellin transforms span the 
space of the functions in Mellin space. After that one only has to solve a 
set of linear equations to make the inverse transform. In the case of two 
loop moments of structure functions in deep inelastic scattering, the 
results in Mellin space are just $S$-functions of weight 4. Because the 
whole space of such $S$-functions is 54 dimensional (a basis has 54 
elements) one has to find 54 functions in $x$-space that map into the 
Mellin space in a linearly independent way. This does not present too many 
problems. One should of course note that this method depends on having 
routines to do the Mellin transforms automatically.

For higher weights it may not be so easy to find a complete set of 
functions in $x$-space. This can be illustrated by a simple calculation. To 
obtain a complete set of functions in $x$-space for which the Mellin 
transforms cover the natural basis of weight $w$ one needs $2\times 3^{w\minus 
1}$ functions in $x$-space. Because this number can be divided by two (the 
relevant functions are of the types $f(x)/(1\pm x)$) only $3^{w\minus 1}$ 
functions have to be considered. A number of these can be constructed by 
taking products of functions that contribute to lower weights. That leaves 
a number of functions that are new at the given weight. This number 
increases rapidly with the weight. They are $3,8,18,48,116$ for the weights 
$3,4,5,6,7$ respectively. Hence one has to come up with a rather large 
number of new functions when the weight becomes large. Fortunately there is 
a method that will work provided only a numerical answer is needed for any 
value of $x$.

Assume that for a given weight $w$ all necessary functions in $x$-space are 
known. Assume also that the Mellin transform of some $F$ is given by
\begin{eqnarray}
\label{eq:definv}
\int_0^1 dx \ x^n F(x) & = & \frac{S_{\scrarrow{m}}(
				n\plus 1)}{(n\plus 1)^p}
\end{eqnarray}
in which ${\overrightarrow{m}}$ represents any allowable series of the type 
$m_1,\cdots,m_q$ and $p > 0$. For this function $F$ one has
\begin{eqnarray}
\label{eq:plusinc}
\int_0^1 dx \ x^n \frac{F(x)}{1\plus x} & = &
		\sign(n)S_{\minus p,\scrarrow{m}}(n)
		-\sign(n)S_{\minus p,\scrarrow{m}}(\infty) \\
\label{eq:mininc}
\int_0^1 dx \frac{x^n F(x)}{(1\minus x)_+} & = &
		S_{p,\scrarrow{m}}(\infty)
		-S_{p,\scrarrow{m}}(n)-S_1(\infty)F(1)
\end{eqnarray}
In the second expression one can see that $F(1)$ will be nonzero when $p=1$ 
and zero otherwise. This is needed to keep the expression finite. It is 
assumed here that $F(x)$ does not contain a factor $\ln(1\minus x)$ or that 
if it does the other components of $F$ still make that $F(1)=0$. If this is 
not the case there will be more complicated sums of the type of 
appendix~\ref{ap:sumsinf} and the right hand side will have more terms to 
cancel the divergences that are due to $S_{p,\scrarrow{m}}(\infty)$ 
having more than one power of $S_1(\infty)$. Rather than using the sums of 
appendix~\ref{ap:sumsinf} one can also use the algorithms of section 
\ref{sec:mellin} to break down the function $F$ completely.

Considering that a knowledge of all odd or all even moments is sufficient to 
reconstruct $F$ the presence of $\sign(n)$ should not be a problem in the 
end. It does not lead to a doubling of the necessary functions --even 
moments in terms of $N$ correspond to odd moments in terms of $n$--.
One should also observe now that the functions $F(x)/(1\plus x)$ and 
$F(x)/(1\minus x)$ are related to the inverse Mellin transforms of 
$S_{\minus p,\scrarrow{m}}(n)$ and $S_{p,\scrarrow{m}}(n)$ 
respectively. Assume now that the $S_{p,\scrarrow{m}}(n)$ are of 
weight $w$. How does one construct the inverse Mellin transforms of 
functions of weight $w\plus 1$? For this one should have a look at the 
functions
\begin{eqnarray}
F^+(x) & = & \int_0^x dx \frac{F(x)}{1\plus x} \\
F^-(x) & = & \int_0^x dx \frac{F(x)}{1\minus x} \\
F^0(x) & = & \int_x^1 dx \frac{F(x)}{x}
\end{eqnarray}
For these functions one can derive readily by means of partial integration
\begin{eqnarray}
\int_0^1 dx\ x^n F^+(x) & = & -\frac{\sign(n\plus 1)}{n\plus 1}
		S_{\minus p,\scrarrow{m}}(n\plus 1)
		+\frac{1}{n\plus 1}(\sign(n\plus 1)-1)
		S_{\minus p,\scrarrow{m}}(\infty) \\
\int_0^1 dx\ x^n F^-(x) & = &\frac{1}{n\plus 1}
			S_{p,\scrarrow{m}}(n\plus 1) \\
\int_0^1 dx\ x^n F^0(x) & = &
		-\frac{1}{(n\plus 1)^{p+1}}S_{\scrarrow{m}}(n\plus 1)
\end{eqnarray}
With the aid of equations~(\ref{eq:plusinc}) and (\ref{eq:mininc}) one 
derives now the relations
\begin{eqnarray}
\label{eq:plusplus}
\int_0^1 dx\ x^n\frac{F^+(x)}{1\plus x} & = &
			-\sign(n)S_{1,\minus p,\scrarrow{m}}(n)
			+\sign(n)(S_{1,\minus p,\scrarrow{m}}(\infty)
				-S_1(\infty)S_{\minus p,\scrarrow{m}}(\infty)) \nn
			+\sign(n)S_{\minus p,\scrarrow{m}}(\infty)(
				S_1(n)-S_{\minus 1}(n)+S_{\minus 1}(\infty))
				\\
\int_0^1 dx\ x^n\frac{F^-(x)}{1\plus x} & = &
		\sign(n)(S_{\minus 1,p,\scrarrow{m}}(n)
			-S_{\minus 1,p,\scrarrow{m}}(\infty))
				\\
\int_0^1 dx\ x^n\frac{F^0(x)}{1\plus x} & = &
		\sign(n)(-S_{\minus(p\plus 1),\scrarrow{m}}(n)
			+S_{\minus(p\plus 1),\scrarrow{m}}(\infty))
				\\
\int_0^1 dx\frac{x^n F^+(x)}{(1\minus x)_+} & = &
		S_{\minus 1,\minus p,\scrarrow{m}}(n)
		-S_{\minus 1,\minus p,\scrarrow{m}}(\infty) \nn
		+S_{\minus p,\scrarrow{m}}(\infty)(S_{\minus 1}(\infty)
			-S_{\minus 1}(n)+S_1(n))
				\\
\int_0^1 dx\frac{x^n F^-(x)}{(1\minus x)_+} & = &
		-S_{1,p,\scrarrow{m}}(n)
		+S_{1,p,\scrarrow{m}}(\infty)-S_1(\infty)
			S_{p,\scrarrow{m}}(\infty)
				\\
\label{eq:zerominus}
\int_0^1 dx\frac{x^n F^0(x)}{(1\minus x)_+} & = &
		S_{(p\plus 1),\scrarrow{m}}(n)
		-S_{(p\plus 1),\scrarrow{m}}(\infty)
\end{eqnarray}
In these expressions is assumed that $F(x)$ contains no factors 
$\ln(1\minus x)$. In that case it is not not difficult to see that all 
divergences cancel. When there are factors $\ln(1\minus x)$ the 
expressions become a bit more complicated in the constant terms in order to 
obtain a complete cancellation of the divergences. The first terms of the 
right hand side expressions form indeed a complete set of $S$-functions of 
weight $w\plus 1$ when all possible values of $p$ and all possible 
$S$-functions in equation~(\ref{eq:definv}) are considered. Because all 
other terms in the right hand side expressions are of a lower weight in 
terms of the argument $n$, their inverse Mellin transforms are supposed to 
be known and hence all inverse Mellin transforms of weight $w\plus 1$ can 
be constructed. If the integrals in the definitions of $F^+$, $F^-$ and 
$F^0$ cannot be solved analytically, one can still obtain their values 
numerically by standard integration techniques. If one has to go more than 
one weight beyond what is analytically possible, one obtains multiple 
integrals. Many of these can of course be simplified by partial 
integrations as can be seen in the following formula:
\begin{eqnarray}
	F^{++}(x) & = & \int_0^x\frac{dx}{1\plus x}
					\int_0^x\frac{dx}{1\plus x}F(x) \\
			  & = & \ln(1\plus x)\int_0^x\frac{dx}{1\plus x}F(x)
				   -\int_0^x \frac{\ln(1\plus x)\ F(x)}{1\plus x} dx
\end{eqnarray}

At this point it seems best to give some examples. First look at the 
constant function in Mellin space. It is the only function with weight zero 
and its inverse Mellin transform is $\delta(1\minus x)$. Here $\delta(x)$ 
is the Dirac delta function. Hence the inverse Mellin transforms for 
functions with weight one are:
\begin{eqnarray}
	\sign(n)S_{\minus 1}(n) & \rightarrow & \frac{1}{1\plus x}
					+ \sign(n)\ln(2)\ \delta(1\minus x) \\
	S_1(n) & \rightarrow & -\frac{1}{1\minus x}
\end{eqnarray}
The factor $\sign(n)$ in the right hand side indicates that the 
reconstruction from the even moments different from the reconstruction from 
the odd moments. This means that if the moments are obtained for even 
values of $N$ (which means odd values for $n$) one should treat the terms 
in $S_{\scrarrow{m}}(n)$ differently from the terms in 
$S_{\scrarrow{m}}(n\plus 1)$.

Next are the functions with weight 2. The only function with weight one 
that can occur in equation (\ref{eq:definv}) is $1/(n\plus 1)$ and its 
inverse Mellin transform is given by $F(x)=1$. From this one can construct 
$F^+(x) = \ln(1\plus x)$, $F^-(x) = -\ln(1\minus x)$ and $F^0(x) = \ln(x)$.
One can now work out the equations (\ref{eq:plusplus} -
\ref{eq:zerominus}) to obtain the inverse Mellin transforms for the weight 
two functions.

For the weight three functions one obtains dilogarithms with the arguments 
$x$, $\minus x$ and $(1\plus x)/2$ as new objects. For the weight four 
functions the functions $F^\pm(x)$ and $F^0(x)$ can have trilogs with the 
arguments $x$, $\minus x$, $(1\plus x)/2$, $1/(1\plus x)$, $1\minus x$, 
$2x/(1\plus x)$, $(1\minus x)/(1\plus x)$ and $\minus(1\minus x)/(1\plus 
x)$. Of course one may choose a different representation in which the 
function $S_{1,2}(x)$ plays a r\^ole (see references \cite{DevotoDuke} and 
\cite{Willy}).

There is one more important observation to be made. The expressions 
(\ref{eq:plusplus}-\ref{eq:zerominus}) have just a single $S$-function of 
weight $w\plus 1$ in the right hand side. This means that one can obtain 
the inverse Mellin transforms of the various $S$-functions without having 
to solve sets of equations. One only has to move terms from the right hand 
side and put their inverse Mellin transform (which is much simpler) into 
the various $F$-functions. This can be done systematically and it can be 
checked by the Mellin transformation program. The approach of 
looking for which functions can occur and then making their Mellin 
transform and inverting the set of equations would lead to very complicated 
sets of equations when the weights become large. 
Hence the interesting functions are more or less the ones 
that have been built up from the original weight one functions by composing 
higher and higher integrals like $F^{+-0++0}(x)$ etc. without writing the 
result in terms of individual polylogarithms.


\section{Some applications}

The values at infinity of the previous section have some 
rather relevant applications for certain classes of integrals. This can 
best be illustrated with some examples. The following integral would under 
normal circumstances be rather difficult, but with all the above tools it 
becomes rather trivial:
\begin{eqnarray}
	\int_0^1 dx\frac{\ln(x)\ \ln^2(1\minus x)\ \ln(1\plus x)}{x} & = &
		-\sum_{i=0}^\infty\frac{1}{i\plus 1}
			\int_0^1 dx\ x^i\ \ln(x)\ \ln(1\minus x)\ \ln(1\plus x)
\end{eqnarray}
The integral is just one of the Mellin transformations, and hence the 
program will handle it. The sum is of the same type as all other sums in 
the Mellin transformation and hence will be done also by the program. In 
the end the answer is expressed in terms of $S$-functions at infinity which 
are maximally of weight 5 and hence they can be substituted from the tables. 
The final result is:
\begin{eqnarray}
	\int_0^1 dx\frac{\ln(x)\ \ln^2(1\minus x)\ \ln(1\plus x)}{x} & = &
       - \frac{3}{8}\zeta_2\zeta_3
       - \frac{2}{3}\zeta_2\ln^3(2)
       + \frac{7}{4}\zeta_3\ln^2(2)
       - \frac{7}{2}\zeta_5
		\nn
       + 4\ln(2)\ \Li_4(1/2)
       + \frac{2}{15}\ln^5(2)
       + 4\ \Li_5\ (1/2)
\end{eqnarray}
Similarly one obtains
\begin{eqnarray}
	\int_0^1 dx\frac{\ln(x)\ \ln^2(1\minus x)\ \ln^2(1\plus x)}{x}
		& = &
       - \frac{1}{2}\zeta_2\ln^4(2)
       - \frac{129}{140}\zeta_2^3
       + \frac{7}{6}\zeta_3\ln^3(2)
		\nn
       - \frac{37}{16}\zeta_3^2
       - \frac{31}{8}\zeta_5\ln(2)
       + 8\ln(2)\ \Li_5(1/2)
       + 4\ln^2(2)\ \Li_4(1/2)
		\nn
       + \frac{1}{9}\ln^6(2)
       + 8\ \Li_6(1/2)
       + 2\ S_{\minus 5,\minus 1}(\infty)
\end{eqnarray}
and the even more difficult integral
\begin{eqnarray}
	\int_0^1 dx\frac{\ln(1\minus x) \Li_2(\frac{1\plus x}{2})
		\Li_3(\frac{1\minus x}{1\plus x})}{1\plus x} & = &
       - \frac{7}{4}\zeta_2\zeta_3\ln^2(2)
       - \frac{5673}{448}\zeta_2\zeta_5
       - 5\zeta_2\ln(2)\ \Li_4(1/2)
       - \frac{17}{120}\zeta_2\ln^5(2)
		\nn
       - 5\zeta_2\Li_5(1/2)
       + \frac{1517}{1120}\zeta_2^2\zeta_3
       + \frac{5}{6}\zeta_2^2\ln^3(2)
       - \frac{1}{84}\zeta_2^3\ln(2)
		\nn
       - \frac{7}{96}\zeta_3\ln^4(2)
       - \frac{3}{4}\zeta_3\Li_4(1/2)
       - \frac{1563}{448}\zeta_3^2\ln(2)
       - \frac{93}{32}\zeta_5\ln^2(2)
		\nn
       + \frac{74415}{1792}\zeta_7
       - 18\ln(2)\ \Li_6(1/2)
       - \frac{43}{14}\ln(2)\ S_{\minus 5,\minus 1}(\infty)
		\nn
       - 6\ln^2(2)\ \Li_5(1/2)
       - \ln^3(2)\ \Li_4(1/2)
       - \frac{1}{84}\ln^7(2)
		\nn
       - 24\ \Li_7(1/2)
       - \frac{45}{7}S_{\minus 5,1,1}(\infty)
       + \frac{32}{7}S_{5,\minus 1,\minus 1}(\infty)
\end{eqnarray}
As one can see, this technique allows the evaluation of whole classes of 
integrals that go considerably beyond the integrals in ref~\cite{DevotoDuke}.

Another application of the techniques of the previous sections concerns the 
evaluation of certain classes of Feynman diagrams. When one tries to 
evaluate moments of structure functions in perturbative QCD one has Feynman 
diagrams which contain the momenta $P$ and $Q$. Assuming that the partons 
are massless one has that $P^2 = 0$ and because all dimensions are 
pulled out of the integral in the form of powers of $Q^2$, there is 
only a single dimensionless kinematic variable left which is $x = 2P\mydot 
Q/Q^2$. The power series expansion in terms of $P$ before integration 
corresponds to the expansion in terms of Mellin moments of the complete 
function after integration. The complete functions have been calculated for 
the two loop level~\cite{Willy} but for the three loop level the 
calculation could only be done for a small number of fixed moments $2,4,6,
8$ and in one case also $10$~\cite{LNRV}. To evaluate all these moments 
requires that the expansion in $P$ should be in terms of a symbolic power 
$N$. This will introduce sums and these sums will be expressed in terms of 
harmonic series. After all integrals have been done all attention has to be 
focussed on the summations and it is actually for this purpose that the 
program SUMMER has been developed. By now a general two loop program has 
been constructed~\cite{twoloop} and studies are on their way to create a 
three loop program. It should be noted that in the two loop program no 
series at infinity can occur. This puts a restriction on the functions that 
can occur in $x$-space. They have to appear in such linear combinations 
that all the constants (with the exception of $\zeta_3$ which comes from 
expansions of the $\Gamma$-function) should cancel in the Mellin transform.


\section{Conclusions}

The algorithms presented in this paper provide a base for working with the 
sums that can occur in many types of calculations, one of which is the 
evaluation of Feynman diagrams in deep inelastic scattering. Additionally 
they allow the analytic evaluation of whole classes of integrals. The 
problem of the Mellin transforms of whole categories of functions has been 
solved, and a numerical solution for inverse Mellin transforms has been 
given. Most of the algorithms and tables have been programmed in the 
language of FORM version 3 and are available from the homepage of the 
author (http://norma.nikhef.nl/$\sim$t68/summer).

The author wishes to thank D.A. Broadhurst, T. van Ritbergen and F.J. 
Yndur\'ain for discussions and support during the various phases of this 
project. He is also indebted to S.A. Larin for the suggestion to have a 
look at these sums.

\appendix
\section{Conversion to the Basis}

To convert products of $S$-functions with an identical last argument to the 
basis of single higher $S$-functions one can use a recursion. If one starts 
with the functions $S^{(1)}$ and $S^{(2)}$ and accumulates the results 
into the function $S^{(3)}$ the recursion reads:
\begin{eqnarray}
    S^{(1)}_{m_1j_1\cdots j_r}(n)
        S^{(2)}_{m_2p_1\cdots p_s}(n)S^{(3)}_{q_1\cdots q_t}(n)
        & \rightarrow &
    S^{(1)}_{m_1j_1\cdots j_r}(n)
        S^{(2)}_{p_1\cdots p_s}(n)S^{(3)}_{q_1\cdots q_tm_2}(n) \nn
    +S^{(1)}_{j_1\cdots j_r}(n)
        S^{(2)}_{m_2p_1\cdots p_s}(n)S^{(3)}_{q_1\cdots q_tm_1}(n) \nn
    -S^{(1)}_{j_1\cdots j_r}(n)
        S^{(2)}_{p_1\cdots p_s}(n)S^{(3)}_{q_1\cdots q_t(m_1 \& m_2)}(n)
\end{eqnarray}
The recursion starts with $S^{(3)}(n) = 1$ and the recursion terminates 
when either $S^{(1)}(n)$ or $S^{(2)}(n)$ has no more indices and hence can 
be replaced by $1$ after which
\begin{eqnarray}
    S^{(a)}_{j_1\cdots j_r}(n)
        S^{(3)}_{q_1\cdots q_t}(n)
        & \rightarrow & S_{q_1\cdots q_tj_1\cdots j_r}(n)
\end{eqnarray}
with $a = 1,2$. Because this is a direct construction of the result, it is 
rather fast. It can be implemented in the language of FORM (version 3 or 
higher) very efficiently:
\begin{verbatim}
  repeat;
    id,once,S(R(?a),n?)*S(R(?b),n?) = SS(R(?a),R,R(?b),n);
    repeat id SS(R(m1?,?a),R(?b),R(m2?,?c),n?) =
                +SS(R(m1,?a),R(?b,m2),R(?c),n)
                +SS(R(?a),R(?b,m1),R(m2,?c),n)
                -SS(R(?a),R(?b,m1*sig_(m2)+m2*sig_(m1)),R(?c),n);
    id,SS(R(?a),R(?b),R(?c),n?) = S(R(?b,?a,?c),n);
  endrepeat;
\end{verbatim}
Note that the function \verb:SS: carries the indices of $S^{(1)}$, $S^{(3)}$
and  $S^{(2)}$ in this order. The function \verb:sig_: returns the sign of 
its argument. Hence the expression that uses this function is one way 
of writing the pseudo addition $\&$.

The above code has been made into a FORM procedure. 
A rather nontrivial test program could be:
\begin{verbatim}
  #-
  #include nndecl.h
  .global
  L F = S(R(1,1,1,1,1),n)*S(R(-1,-1,-1,-1,-1),n);
  #call basis(S);
  .end
\end{verbatim}
It gives the result
\begin{verbatim}
  Time =       0.64 sec    Generated terms =       1683
                  F        Terms in output =       1683
                           Bytes used      =      85104
\end{verbatim}
The run was made on a Pentium Pro 200 chip running the NeXTstep operating 
system. As one can see, these expressions can become rather complicated. On 
the other hand, weight 10 functions are of course not trivial. It should be 
noted that it is relatively easy to test routines like the one above. 
One can try them out for any functions and any values of the argument and 
evaluate the corresponding harmonic series into a rational number and see 
that they are identical.

\section{Conjugations}

\label{ap:conjugations}
For the conjugations one should consider only $S$-functions with positive 
indices. The conjugation is defined with the sum
\begin{equation}
\label{eq:conjugate}
    (f(n))^C = -\sum_{i=1}^n \sign(i)\binom(n,i) f(i)
\end{equation}
That this is a conjugation can be shown easily by applying it twice. 
This gives the original function. For the function $f$ one can use 
$S$-functions or the combination of an $S$-function and a negative power of 
the argument of the $S$-function as in $S_{j_1\cdots j_r}(n)/n^k$. For 
these functions one has:

\noindent {\sl Theorem:}
The conjugate function of an element of the natural basis with 
only positive indices is a single $S$-function of a lower weight with only 
positive indices, combined with enough negative powers of its argument to 
give the complete term the same extended weight as the original function.

\noindent {\sl Proof:}
First look at the weights one and two:
\begin{eqnarray}
    (S_1(n))^C & = & 1/n \\
    (S_2(n))^C & = & S_1(n)/n \\
    (S_{1,1}(n))^C & = & 1/n^2
\end{eqnarray}
They clearly fulfill the theorem. Then write
\begin{equation}
    (S_{mj_1\cdots j_r}(n))^C =
                \frac{1}{n}(S_{j_1\cdots j_r}(n)/n^{m\minus 1})^C
\end{equation}
This identity can be obtained by writing the outermost sum and then 
exchanging it with the sum of the conjugation.
Assume now that the theorem holds for all functions with a lower weight. 
There are two cases: $m=1$ and $m>1$. When $m=1$ the problem has been 
reduced to the same problem of finding the conjugate but now for a function 
with a lower extended weight. Hence, if the theorem holds for all simpler 
functions it holds also at the current weight. For $m>1$ the 
conjugate of $S_{j_1\cdots j_r}(n)/n^{m-1}$ must be a single harmonic 
function of weight $m-1$. This can be seen when one realizes that for each 
extended weight there are as many functions with their `proper' weight equal 
to this extended weight as with their `proper' weight less than the 
extended weight. Hence the function must have a conjugate that is a single 
$S$-function. Together with the fact that two conjugations give the 
original function, and the fact that all $S$-functions of a given weight are 
linearly independent this completes the proof of the theorem.

Next is the derivation of an algorithm to find the conjugate of 
$S_{j_1\cdots j_r}(n)/n^{m-1}$. One way would be to successively build up 
the algorithm by first deriving all conjugates up to a given weight. After 
that one can obtain the needed conjugates by reading the formulae 
backwards. This is not very elegant. For a more direct way one can define 
the concept of the associate function.
\begin{eqnarray}
    S_{j_1\cdots j_r}^A(n) & = & \sum_{i=1}^n(S_{j_1\cdots j_r}(i))^C
\end{eqnarray}
Note that because $X = S_{j_1\cdots j_r}$ is an element of the basis, 
$(X(i))^C$ contains powers of $1/i$ and the sum gives again a single 
harmonic function of the same weight as $X$. It is rather easy to prove that 
$(X^A)^A = X$. The task of finding the conjugate can now be reduced to the 
task of finding the associate function. If this associate function can be 
written as $S_{mj_1\cdots j_r}(n)$ the conjugate will be 
$S_{j_1\cdots j_r}(n)/n^m$. Similarly 
a function in combination with negative powers of $n$ can be rewritten as a 
sum ($S_{j_1\cdots j_r}(n)/n^m\rightarrow S_{mj_1\cdots j_r}(n)$), 
and then the associate function of 
this function will be the needed conjugate function.

The associate function can be found by construction. Assume that
$(S_{j_1\cdots j_r}(n))^A = S_{mp_1\cdots p_s}(n)$. Then
\begin{eqnarray}
    (S_{1j_1\cdots j_r}(n))^A & = &
        \sum_{i=1}^n(S_{1j_1\cdots j_r}(i))^C \nonumber \\
                & = & \sum_{i=1}^n\frac{1}{i}(S_{j_1\cdots j_r}(i))^C \nonumber \\
                & = & \sum_{i=1}^n\frac{1}{i}
                    \frac{S_{p_1\cdots p_s}(i)}{i^m} \nonumber \\
                & = & S_{(m\plus 1)p_1\cdots p_s}(n)
\end{eqnarray}
and similarly for $k>1$:
\begin{eqnarray}
    (S_{kj_1\cdots j_r}(n))^A & = &
        \sum_{i=1}^n(S_{kj_1\cdots j_r}(i))^C \nonumber \\
& = & \sum_{i=1}^n\frac{1}{i}(\frac{S_{j_1\cdots j_r}(n)(i)}{i^{k\minus 1}})^C 
                                        \nonumber \\
                & = & \sum_{i=1}^n\frac{1}{i}(S_{(k\minus 1)j_1\cdots 
                        j_r}(i))^A \nonumber \\
                & = & S_{1q_1\cdots q_t}(n)
\end{eqnarray}
with $(S_{(k\minus 1)j_1\cdots j_r}(n))^A = S_{q_1\cdots q_t}(n)$.
Considering that $(S_1(n))^A = S_1(n)$ associate functions to any weight 
can now be constructed. This algorithm is also easy to implement in a 
program like FORM. 

\section{Sums involving $n-i$}

\label{ap:nmini}
In this appendix sums of the type
\begin{equation}
    \sum_{i=1}^{n\minus 1} \frac{S_{p_1\cdots p_s}(n\minus i)
            S_{q_1\cdots q_t}(i)}{i^k}
\end{equation}
will be considered. 
It is impossible to combine the sums to a single basis 
element. Hence a different method is called for. Assume first that $k > 
0$ and $m > 0$ (below). Writing out the outermost 
sum of the $S$-function with the argument $n\minus i$ leads to
\begin{eqnarray}
    \sum_{i=1}^{n\minus 1} \frac{S_{mp_1\cdots p_s}(n\minus i)
            S_{q_1\cdots q_t}(i)}{i^k} & = &
    \sum_{i=1}^{n} \sum_{j=1}^{n\minus i}
        \frac{S_{p_1\cdots p_s}(j)}{j^m}
            \frac{S_{q_1\cdots q_t}(i)}{i^k} \nonumber \\ & = &
    \sum_{i=1}^n \sum_{j=i\plus 1}^{n}
        \frac{S_{p_1\cdots p_s}(j\minus i)}{(j\minus i)^m}
            \frac{S_{q_1\cdots q_t}(i)}{i^k} \nonumber \\ & = &
    \sum_{j=1}^n \sum_{i=1}^{j\minus 1}
        \frac{S_{p_1\cdots p_s}(j\minus i)}{(j\minus i)^m}
            \frac{S_{q_1\cdots q_t}(i)}{i^k} \nonumber \\ & = &
    \sum_{i=1}^n \sum_{j=1}^{i\minus 1}
        S_{p_1\cdots p_s}(i\minus j) S_{q_1\cdots q_t}(j)
            \frac{1}{(i\minus j)^m j^k}
\end{eqnarray}
Partial fractioning of the denominators gives sums in which 
the denominator is a power $k' \le k$ of $j$ and sums in which the 
denominator is a power $m' \le m$ of $i\minus j$. These last sums can be 
done immediately by reverting the direction of summation. Hence:
\begin{eqnarray}
    \sum_{i=1}^{n\minus 1} \frac{S_{mp_1\cdots p_s}(n\minus i)
            S_{q_1\cdots q_t}(i)}{i^k} & = &
    \sum_{a=1}^k \binom(k\plus m\minus 1\minus a,m\minus 1)
    \sum_{i=1}^n \sum_{j=1}^{i\minus 1}
        S_{p_1\cdots p_s}(i\minus j) S_{q_1\cdots q_t}(j)
            \frac{1}{i^{m\plus k\minus a} j^a}  \nn
    +\sum_{a=1}^m \binom(k\plus m\minus 1\minus a,k\minus 1)
    \sum_{i=1}^n \sum_{j=1}^{i\minus 1}
        S_{p_1\cdots p_s}(j) S_{q_1\cdots q_t}(i\minus j)
            \frac{1}{i^{m\plus k\minus a} j^a}
\end{eqnarray}
Now the innermost sum is of a simpler type. Hence eventually one can do 
this sum, and after that all remaining sums are rather simple. Therefore 
this defines a useful recursion. When a negative value of $m$ or a factor 
$\sign(i)$ is involved things are only marginally more complicated. This 
algorithm has been programmed in FORM and carries the name sumnmii. An 
example of its application is 
\begin{verbatim}
  #-
  #include nndecl.h
  .global
  L F = sum(j,1,n-1)*S(R(1,2,1),n-j)*S(R(-2,-1,-2),j)/j^2;
  #call sumnmii()
  .end
\end{verbatim}
with the output
\begin{verbatim}
  Time =       0.28 sec    Generated terms =        478
                  F        Terms in output =        208
                           Bytes used      =       9148
\end{verbatim}
which are all terms with a single function of weight 11.

The algorithm for doing the sums of the type
\begin{equation}
    G_{p_1\cdots p_s}^{q_1\cdots q_t}(k,n)
	 = -\sum_{i=1}^{n\minus 1}\sign(i)\binom(n,i)
		 \frac{S_{p_1\cdots p_s}(n\minus i)
            S_{q_1\cdots q_t}(i)}{i^k}
\end{equation}
is more complicated. First one has to assume that all $p_j$ and $q_j$ are 
positive. Assuming also that $k \ge 0$, one can derive
\begin{eqnarray}
  G_{mp_1\cdots p_s}^{q_1\cdots q_t}(k,n) & = &
		G_{mp_1\cdots p_s}^{q_1\cdots q_t}(k,n\minus 1)
		+\frac{1}{n}G_{mp_1\cdots p_s}^{q_1\cdots q_t}(k-1,n) \nn
	+\frac{1}{n}\sum_{i=1}^n\sign(i\plus 1)\binom(n,i)
		\frac{S_{p_1\cdots p_s}(n\minus i)S_{q_1\cdots q_t}(i)}
			{(n\minus i)^{m\minus} i^k}
\end{eqnarray}
Because the weight of the $G$-function in the second term is one less, and 
because one can partial fraction the last term in the end all terms 
have a sum over a combination with a lower weight. This means that one can 
use this equation for a recursion, provided one knows how to deal with the 
case $k = 0$ which is not handled by the above equation. For $k = 0$ one 
obtains after some algebra
\begin{eqnarray}
  G_{m_1p_1\cdots p_s}^{m_2q_1\cdots q_t}(0,n) & = &
	+\frac{1}{n}\sum_{i=1}^n\sign(i\plus 1)\binom(n,i)
		\frac{S_{m_1p_1\cdots p_s}(n\minus i)S_{q_1\cdots q_t}(i)}
			{i^{m_2\minus 1}} \nn
	+\frac{1}{n}\sum_{i=1}^n\sign(i\plus 1)\binom(n,i)
		\frac{S_{p_1\cdots p_s}(n\minus i)S_{m_2q_1\cdots q_t}(i)}
			{(n\minus i)^{m_1\minus 1}}
\end{eqnarray}
Hence also here the weight has been decreased and one can use it for a 
recursion. The final expression for $G$ can be obtained by an extra sum, 
because $G(k,0) = 0$ for all indices and one obtains an 
expression for $G(k,n)-G(k,n\minus 1)$. One should also realize that in 
some cases it is necessary to change the direction of the sum ($i 
\rightarrow n\minus i$) which will introduce terms of the type $\sign(n)$ 
and hence this last sum can give $S$-functions with a negative index.

The routine that implements these algorithms (sumnmic) is a bit lengthy. 
A test run gives
\begin{verbatim}
  #-
  #include nndecl.h
  .global
  L  F = sum(j,1,n)*sign(j)*bino(n,j)*S(R(1,2,1),n-j)*S(R(2,1,2),j)/j^2;
  #call sumnmic()
  .end
\end{verbatim}
with the result
\begin{verbatim}
  Time =       0.36 sec    Generated terms =        238
                  F        Terms in output =        131
                           Bytes used      =       6478
\end{verbatim}
and a simpler example gives
\begin{verbatim}
  #-
  #include nndecl.h
  .global
  L F = sum(j,1,n)*sign(j)*bino(n,j)*S(R(2),n-j)/j;
  #call sumnmic()
  Print;
  .end

     F =
         - S(R(-3),n) - 2*S(R(-2,1),n) - S(R(1,2),n) - S(R(2,1),n)
         - S(R(3),n);
\end{verbatim}

\section{Some sums to infinity}

\label{ap:sumsinf}
There are special classes of sums for which the upper bound is infinity. A 
number of them can be evaluated to any level of complexity. Consider for 
instance the following sum (with $p_1 > 0$; negative values just give extra 
powers of $-1$):
\begin{equation}
	F(m) = \sum_{j=1}^\infty \frac{S_{p_1\cdots p_r}(j\plus m)}{j^k}
\end{equation}
Such sums can be evaluated by setting up a sum over m:
\begin{eqnarray}
	F(m) & = & F(m\minus 1) +
		 \sum_{j=1}^\infty \frac{S_{p_2\cdots p_r}(j\plus m)}{(m\plus 
			j)^{p_1}j^k} \nonumber \\ & = &
		 F(m\minus 1) +
		\sum_{i=1}^{p_1}\binom(p_1\plus k \minus 1\minus i,k\minus 1)
			\frac{1}{m^{p_1\plus k\minus i}}
				(S_{i,p_2,\cdots,p_r}(\infty)-S_{i,p_2,\cdots,p_r}(m)) \nn
		+\sum_{i=1}^{k}\binom(p_1\plus k \minus 1\minus i,p_1\minus 1)
			\frac{\sign(p_1\plus k\minus i)}{m^{p_1\plus k\minus i}}
			\sum_{j=1}^\infty\frac{S_{p_2\cdots p_r}(j\plus m)}{j^i}
\end{eqnarray}
The sum in the last term is of the same type as the original sum, but it is 
of a simpler nature. The sum over $i$ can just be worked out, because $k$ 
and $p_1$ are just numbers. Hence this defines a recursion which can be 
worked out, if not by hand, then by computer. In the end one obtains an 
expression for $F(m) - F(m\minus 1)$ which can be summed:
\begin{eqnarray}
	F(m) & = & F(0) + \sum_{i=1}^m ( F(i) - F(i\minus 1) ) \nne
		S_{kp_1\cdots p_r}(\infty)
		+ \sum_{i=1}^m ( F(i) - F(i\minus 1) )
\end{eqnarray}

Similarly one can consider sums of the type
\begin{equation}
	F(m) = \sum_{j=1}^\infty \frac{S_{p_1\cdots p_r}(j)}{(j\plus m)^k}
\end{equation}
The technique to construct a recursive solution for these sums is similar. 
One can study the function
\begin{eqnarray}
	F(m)-F(m\minus 1) & = &
		\sum_{j=1}^\infty \frac{S_{p_1\cdots p_r}(j)}{(j\plus m)^k}
		-\sum_{j=0}^\infty \frac{S_{p_1\cdots p_r}(j\plus 1)}{(j\plus m)^k}
		\nonumber \\ & = &
		-\frac{S_{p_1\cdots p_r}(1)}{m^k}
		-\sum_{j=0}^\infty \frac{S_{p_2\cdots p_r}(j\plus 1)}{
				(j\plus 1)^{p_1}(j\plus m)^k}
		\nonumber \\ & = &
		-\frac{S_{p_1\cdots p_r}(1)}{m^k}
		-\sum_{i=1}^{p_1}\binom(p_1\plus k \minus 1\minus i,k\minus 1)
			\frac{\sign(p_1\plus k\minus i)}{(m\minus 1)^{p_1\plus k\minus i}}
				S_{i,p_2,\cdots,p_r}(\infty) \nn
		-\sum_{i=1}^{k}\binom(p_1\plus k \minus 1\minus i,p_1\minus 1)
			\frac{1}{(m\minus 1)^{p_1\plus k\minus i}}
		\sum_{j=0}^\infty \frac{S_{p_2\cdots p_r}(j\plus 1)}{(j\plus m)^i}
\end{eqnarray}
and again the last term is of a simpler nature. Hence there is a useful 
recursion and these sums can be solved.

In both cases there will be some $S$-functions in the answer that have the 
argument infinity. These should not present any special problems as they 
have been discussed before.


\section{Miscellaneous Sums}

\label{ap:miscel}
In this section some sums are given that can be worked out to any level of 
complexity, but they are not representing whole classes. Neither is there 
any proof for the algorithms. The algorithms presented have just been 
checked up to some rather large values of the parameters.

The sums that are treated here involve two binomial coefficients. 
There are quite a few of these sums in appendix \ref{ap:sumtables}, but 
here are the ones that can be done to any order. The 
first relation that is needed is:
\begin{equation}
	\sum_{j=0}^m\sign(j)\binom(m\plus i\plus j,i\plus j)
	\binom(m\plus 2i\plus j,m\plus i\plus j) =
		\sign(m)\binom(m\plus i,i)\binom(m\plus 2i,i)
\end{equation}
Taking $m=n\minus i$ leads to:
\begin{eqnarray}
	\sum_{j=0}^n \sign(j)\binom(n,j)\binom(n\plus j,m\plus j)f^C(m\plus j) & = &
	-\sign(n\plus m)\sum_{j=0}^n \sign(j)\binom(n,j)\binom(n\plus j,m\plus j)f(m\plus j)
\end{eqnarray}
for $0 \le m \le n$. Here $f^C$ indicates the conjugation of appendix 
\ref{ap:conjugations}. This is a rather useful identity as it divides the 
necessary amount of work by two. Alternatively it may even make terms 
cancel and hence make further evaluation unnecessary.

A new function is needed to keep the notation short:
\begin{equation}
	U_k(n,m) = S_k(n\plus m) - \sign(k)S_k(n\minus m) - S_k(m\minus 1)
\end{equation}
for $k,n \ge 0$ and $m > 0$. $U_0(n,m)$ is defined to be one.

One of the ways the harmonic series can be introduced in many calculations 
is by expansion of the $\Gamma$-function. At the negative side its expansion 
is:
\begin{eqnarray}
	\Gamma(-n+\epsilon) & = & \frac{\sign(n)}{\epsilon n!}
		\Gamma(1+\epsilon)(1+S_1(n)\epsilon
	+S_{1,1}(n)\epsilon^2
	+ S_{1,1,1}(n)\epsilon^3
	+ S_{1,1,1,1}(n)\epsilon^4
	+ \cdots )
\end{eqnarray}
Actually these special harmonic series can be written as a sum of terms 
that contain only products of harmonic series with a single sum as in:
\begin{eqnarray}
	2S_{1,1}(n) & = & (S_1(n))^2+S_2(n) \\
	6S_{1,1,1}(n) & = & (S_1(n))^3 + 3S_1(n)S_2(n)+2S_3(n) \\
	24S_{1,1,1,1}(n) & = &
		(S_1(n))^4+6(S_1(n))^2S_2(n)+8S_1(n)S_3(n)+3(S_2(n))^2+6S_4(n)
\end{eqnarray}
Notice that the factors are related to the cycle structure of the 
permutation group. One can define the higher $U$ functions by analogy:
\begin{eqnarray}
	2U_{1,1}(n,m) & = & (U_1(n,m))^2+U_2(n,m) \\
	6U_{1,1,1}(n,m) & = & (U_1(n,m))^3 + 3U_1(n,m)U_2(n,m)+2U_3(n,m) \\
	24U_{1,1,1,1}(n,m) & = &
		(U_1(n,m))^4+6(U_1(n,m))^2U_2(n,m)+8U_1(n,m)U_3(n,m) \nn
			+3(U_2(n,m))^2+6U_4(n,m)
\end{eqnarray}
With these definitions one can write ($0 < m \le n$):
\begin{eqnarray}
	\sum_{j=0}^n\sign(j)\binom(n,j)\binom(n\plus j,m\plus j)\frac{1}{(m\plus 
		j)^1} & = & \frac{n!\ (m\minus 1)!}{(n\plus m)!} \\
	\sum_{j=0}^n\sign(j)\binom(n,j)\binom(n\plus j,m\plus j)\frac{1}{(m\plus 
		j)^2} & = & \frac{n!\ (m\minus 1)!}{(n\plus m)!}\ U_1(n,m) \\
	\sum_{j=0}^n\sign(j)\binom(n,j)\binom(n\plus j,m\plus j)\frac{1}{(m\plus 
		j)^3} & = & \frac{n!\ (m\minus 1)!}{(n\plus m)!}\ U_{1,1}(n,m) \\
	\sum_{j=0}^n\sign(j)\binom(n,j)\binom(n\plus j,m\plus j)\frac{1}{(m\plus 
		j)^4} & = & \frac{n!\ (m\minus 1)!}{(n\plus m)!}\ U_{1,1,1}(n,m)
\end{eqnarray}
etc. In the case that $m$ is zero there are different expressions:
\begin{eqnarray}
	\sum_{j=1}^n\sign(j)\binom(n,j)\binom(n\plus j,j)\frac{1}{j}
	& = & -2S_1(n) \\
	\sum_{j=1}^n\sign(j)\binom(n,j)\binom(n\plus j,j)\frac{1}{j^2}
	& = & -4S_{1,1}(n)+2S_2(n) \\
	\sum_{j=1}^n\sign(j)\binom(n,j)\binom(n\plus j,j)\frac{1}{j^3}
	& = & -8S_{1,1,1}(n)+4S_{1,2}(n)+4S_{2,1}(n)-2S_3(n) \\
	\sum_{j=1}^n\sign(j)\binom(n,j)\binom(n\plus j,j)\frac{1}{j^4}
	& = & -16S_{1,1,1,1}(n)+8(S_{1,1,2}(n)+S_{1,2,1}(n)+S_{2,1,1}(n))
	\nonumber \\ && -4(S_{1,3}(n)+S_{2,2}(n)+S_{3,1}(n))+2S_4(n)
\end{eqnarray}
and the pattern should be clear: For $1/j^k$ there will be all functions with 
weight $k$. The ones with $m$ nested sums have a coefficient $-\sign(k\minus 
m)2^m$. A recipe of a similar type is found for the following sums:
\begin{eqnarray}
	\sum_{j=1}^n\sign(j)\binom(n,j)\binom(n\plus j,j)S_1(j)
	& = & 2\sign(n)S_1(n) \\
	\sum_{j=1}^n\sign(j)\binom(n,j)\binom(n\plus j,j)S_2(j)
	& = & \minus 2\sign(n)S_{\minus 2}(n) \\
	\sum_{j=1}^n\sign(j)\binom(n,j)\binom(n\plus j,j)S_3(j)
	& = & \sign(n)(2S_{\minus 3}(n) \minus 4S_{\minus 2,1}(n)) \\
	\sum_{j=1}^n\sign(j)\binom(n,j)\binom(n\plus j,j)S_4(j)
	& = & \sign(n)(\minus 2S_{\minus 4}(n)
		\plus 4(S_{\minus 3,1}(n)
		\plus S_{\minus 2,2}(n))
	\minus 8S_{\minus 2,1,1}(n))
\end{eqnarray}
and the pattern here is that one should make all higher series that start 
with a negative index that has a value of at most $-2$, after which there 
are only positive indices. All functions are of weight $k$ 
(for $S_k$ inside the sum), and for $m$ nested sums the coefficient is 
$\sign(n\plus k\minus m)2^m$. The exception is $k=1$ but that is because 
$S_1$ is its own associated function and its conjugate is purely of the 
type $1/j^k$.

%
\section{Summation tables}

\label{ap:sumtables}
During the work that inspired this paper quite a few other sums were 
evaluated that are not represented by the above algorithms. Many of these 
sums can only be done for a fixed weight and most of them were not readily 
available in the literature. Hence they are presented here in a number of 
tables, even though eventually many of these sums were not needed in the 
final version of the program. For completeness also a large number of sums 
are presented that are already available in the literature. A number of 
these sums can be derived formally. Some were derived by `guessing' and 
then trying the resulting formula for a large number of values.

In all sums it is assumed that all parameters i,j,k,l,m,n are integers and 
have values $\ge 0$. In some cases the formulae can be extended to 
noninteger values.

It should be noted that all sums that can be handled by the procedures of 
the previous appendices are not in the tables. They would make the tables 
unnecessarily lengthy.

Some formulae that are used very often are presented first:
\begin{eqnarray}
	\sum_{i = 0}^n \frac{(m\plus i)!}{i!} & = &
					\frac{(n\plus m\plus 1)!}{n!(m\plus 1)}
\\
	\sum_{i = 0}^n \frac{(m\plus i)!}{(k\plus i)!} & = &
					\frac{(n\plus m\plus 1)!}{(n\plus k)!(m\plus 1\minus k)}
					-\frac{m!}{(k\minus 1)!(m\plus 1\minus k)}
\\
	\sum_{j=0}^n\sign(j)\binom(n,j) & = & \delta(n)
\\
	\sum_{j=0}^m\sign(j)\binom(n,j) & = & (\minus 1)^m\binom(n\minus 1,m)
\\
	\sum_{j=0}^n\binom(n,j)(m\plus j)!\ (k\plus n\minus j)! & = &
			m!\ k! \frac{(m\plus k\plus n\plus 1)!}{
							(m\plus k\plus 1)!}
\\
	\sum_{j=0}^n\sign(j)\binom(n,j)\frac{(m\plus j)!}{(m\plus k\plus j)!} & = &
			\frac{(n\plus k\minus 1)!}{(k\minus 1)!}\frac{m!}{(m\plus n\plus k)!}
\\
	\sum_{j=0}^n\sign(j)\binom(n,j)\binom(n\plus m\plus k\plus j,m\plus j) & = &
				 (\minus 1)^n\binom(n\plus m\plus k,k)
\end{eqnarray}
The last three formulae can be extended to noninteger values of m and k. 
They can be used occasionally before $\Gamma$-functions are expanded to 
yield harmonic series.

At times some auxiliary functions were needed. They are defined by
\begin{eqnarray}
	A_k(n,m) & = & \sum_{j=1}^n(\minus 1)^j\binom(n+m,j)\frac{1}{j^k} \\
	\Delta(1,1,a,b) & = &
			\sum_{i=a}^b\sum_{j=a}^{a\plus b\minus i}\frac{1}{i\ j} \\
	R_{m,k}(n) & = & \sum_{j=1}^n\frac{S_m(2j)}{j^k}
\end{eqnarray}
Sometimes $\Delta$ is not always the easiest function to manipulate. 
Therefore the function $\Delta'$ is sometimes handy:
\begin{eqnarray}
	\Delta'(1,1,a,b) & = & \frac{1}{2}\theta(b\minus a)(\Delta(1,1,a,b)
			-(S_1(a\minus 1)-S_1(b))^2)
	\nonumber \\ &&
		-\frac{1}{2}\theta(a\minus b\minus 1)\Delta(1,1,b\plus 1,a\minus 1)
\end{eqnarray}
Both functions involve summations over triangles in the two dimensional 
plane. These triangles do not touch the origin.

\subsection{Sums without $\sign(j)$}

First is a number of expressions that are at the lowest level of 
complexity.
\begin{eqnarray}
	A_1(n,1) & = & -S_1(n\plus 1)+\frac{(\minus 1)^n}{n\plus 1}
\\
	A_1(n,2) & = & -S_1(n\plus 2)+\frac{(\minus 1)^n}{(n\plus 1)(n\plus 2)}
				+(\minus 1)^n
\\
	A_1(1,m) & = & -(m\plus 1)
\\
	A_1(2,m) & = & \frac{1}{4}(m\plus 2)(m\minus 3)
\\
	A_1(n,m+1) & = & A_1(n,m)+(\minus 1)^n\binom(n\plus m,n)
				\frac{1}{n\plus m\plus 1}-\frac{1}{n\plus m\plus 1}
\\
	A_1(n+1,m) & = & A_1(n,m+1)-(\minus 1)^n\binom(n\plus m\plus 1,m)
				\frac{1}{n\plus 1}
\\
	A_k(n,m\plus 1) & = & \sum_{i=1}^k(n\plus m\plus 1)^{i-k}A_i(n,m)
			+(\sign(n)\binom(n+m,n)-1)\frac{1}{(n\plus m\plus 1)^k}
\\
	A_k(n,m\plus 1) & = & A_k(n,m) +\sum_{j=1}^n\sign(j)\binom(n\plus m\plus 1,
	j)S_k(j) -\sign(n)\binom(n\plus m,n)S_k(n)
\end{eqnarray}
Next are some miscellaneous sums:
\begin{eqnarray}
	\sum_{j=1}^a\frac{1}{j}S_1(b\plus j) & = & (S_1(a)-S_1(b))S_1(a\plus b)
						+2S_{1,1}(b)-\frac{1}{2}S_2(b)
		\nonumber \\ && + \Delta'(1,1,a,b)
\\
	\sum_{j=0}^m\binom(n\minus 1\plus j,j)\frac{1}{n\plus j} & = &
		\binom(n\plus m,n)\frac{1}{n}
		-(\minus 1)^nA_1(n,m)-(\minus 1)^nS_1(n\plus m)
\\
	\sum_{j=1}^n\frac{S_1(n\plus j)}{j} & = & 2S_{1,1}(n)-\frac{1}{2}S_2(n)
\\
	\sum_{j=1}^n\frac{S_1(j)}{n\plus j} & = &
		S_1(2n)S_1(n\minus 1)-2S_{1,1}(n\minus 1)+\frac{1}{2}S_2(n)
\\
	\sum_{j=1}^n\frac{S_{\minus 1}(n\plus j)}{j} & = &
		S_{\minus 1,\minus 1}(n)+S_{1,\minus 1}(n)-\frac{1}{2}S_2(n)
\\
	\sum_{j=1}^n\frac{S_1(n\plus j)}{j^2} & = &
		S_{1,2}(n)-2S_{2,1}(n)-\frac{1}{2}S_3(n)+2R_{1,2}(n)
\\
	\sum_{j=1}^n\frac{S_{\minus 1}(n\plus j)}{j^2} & = &
      2S_{2,1}(n) - S_{\minus 2,\minus 1}(n) + S_{\minus 1,\minus 2}(n)
		- S_{2,\minus 1}(n)
		 -\frac{1}{2}S_3(n)-2R_{1,2}(n)
\\
	\sum_{j=1}^n\frac{S_{\minus 2}(n\plus j)}{j} & = &
		S_{\minus 2,\minus 1}(n)
		+S_{1,\minus 2}(n)
		+S_{2,\minus 1}(n)
		-S_{2,1}(n)
		-\frac{1}{4}S_3(n)
		+R_{1,2}(n)
\\
	\sum_{j=1}^m\frac{1}{n\plus j\plus 2}S_1(j) & = &
		S_1(m)S_1(n\plus m\plus 2)-\sum_{j=1}^m\frac{1}{j}S_1(n\plus j\plus 1)
\\
	\sum_{j=1}^m\frac{1}{n\plus j\plus 2}S_1(j\plus 1) & = &
			-\sum_{j=1}^m\frac{1}{j\plus 1}S_1(n\plus 1\plus j)
			+S_1(m\plus 1)S_1(n\plus m\plus 2)-S_1(n\plus 2)
\end{eqnarray}
The last two equations are not solving anything, but they are useful in the 
derivation of some of the next sums. First an equation that is like a 
partial integration.
\begin{eqnarray}
	\sum_{j =0}^n\binom(m\plus j,j)f(n\plus 1\minus j) & = &
		\sum_{j =0}^n\binom(m\minus 1\plus j,j)\sum_{i = 1}^{n\plus 1\minus j}f(i)
\end{eqnarray}
It is central in the derivation of the next equations
\begin{eqnarray}
	\sum_{j=0}^n\binom(m\plus j,j)\frac{1}{n\plus 1\minus j} & = &
		\binom(n\plus m\plus 1,n\plus 1)(S_1(n\plus m\plus 1)-S_1(m))
	\\
	\sum_{j=0}^n\binom(m\plus j,j)\frac{1}{(n\plus 1\minus j)^2} & = &
		\binom(n\plus m\plus 1,n\plus 1)(
			S_1(n\plus 1)(S_1(n\plus m\plus 1)-S_1(m\minus 1))
	\nonumber \\ &&
			+2S_2(n\plus m\plus 1)-2S_{1,1}(n\plus m\plus 1)
			+\sum_{i=1}^{m\minus 1}\frac{S_1(n\plus m\plus 1\minus i)}{i})
	\\
	\sum_{j =0}^n\binom(m\plus j,j)S_k(n\plus 1\minus j) & = &
		\frac{n\plus m\plus 2}{m\plus 1}\sum_{j =0}^n\binom(m\plus j\minus 1,j)
				S_k(n\plus 1\minus j)
		\nonumber \\ &&
		-\frac{1}{m\plus 1}\sum_{j =0}^n\binom(m\plus j\minus 1,j)S_{k\minus 1}(n\plus 1\minus j)
	\\
	\sum_{j=0}^n\binom(m\plus j,j)S_1(n\plus 1\minus j) & = &
		\binom(n\plus m\plus 2,n\plus 1)(S_1(n\plus m\plus 2)-S_1(m\plus 1))
	\\
	\sum_{j=0}^n\binom(m\plus j,j)S_{1,1}(n\plus 1\minus j) & = &
		\frac{1}{2}\binom(n\plus m\plus 2,n\plus 1)(
		2S_{1,1}(m\plus 1)-2S_{1,1}(n\plus 1)
		+S_2(n\plus 1)
			\nn
		-\frac{1}{(n\plus m\plus 2)(m\plus 1)}
			-\Delta'(1,1,m,n\plus 1)
			-\Delta'(1,1,m\plus 1,n\plus 1)
			\nn
		+ (2S_1(n\plus m\plus 2)-\frac{1}{n\plus m\plus 2})
			(S_1(n\plus 1)-S_1(m\plus 1))\ )
	\\
	\sum_{j=0}^n\binom(m\plus j,j)S_2(n\plus 1\minus j) & = &
		\binom(n\plus m\plus 2,n\plus 1)(
		S_2(n\plus m\plus 2)-S_{1,1}(n\plus m\plus 2) \nonumber \\ &&
		+S_1(n\plus m\plus 1)S_1(n\plus 1) +\frac{S_1(m)}{n\plus m\plus 2}
		+\frac{1}{2}S_2(n\plus 1) \nonumber \\ &&
		-S_{1,1}(n\plus 1)-\Delta'(1,1,m,n\plus 1)\ )
\end{eqnarray}
Similarly one can derive:
\begin{eqnarray}
	\sum_{j=0}^n\binom(m\plus j,j)S_1(m\plus j) & = &
				\binom(n\plus m\plus 1,n) ( S_1(n\plus m\plus 1)-\frac{1}{m\plus 1})
		\\
	\sum_{j=0}^n\binom(m\plus j,j)S_1^2(m\plus j) & = &
		\binom(n\plus m\plus 1,n)(
			(S_1(n\plus m)-\frac{1}{m\plus 1})(
					\frac{1}{n\plus m\plus 1}-\frac{1}{m\plus 1}) \nn
			+S_{1,1}(n\plus m))
		-\sign(m)\frac{1}{m\plus 1}(S_1(n\plus m)+A_1(m,n))
		\\
	\sum_{j=0}^n\binom(m\plus j,j)S_2(m\plus j) & = &
		\binom(n\plus m\plus 1,n)S_2(n\plus m)
		-\frac{\sign(m)}{m\plus 1}(S_1(n\plus m)+
		A_1(m,n))
\end{eqnarray}
These formulae give also some 'partial integration':
\begin{eqnarray}
	\sum_{j=0}^{n\minus 1}\frac{(m\plus j)!}{j!}S_1(m\plus j)f(n\minus j) & = &
			\sum_{j=0}^{n\minus 1}m\frac{(m\plus j\minus 1)!}{j!}S_1(m\plus j\minus 1)\sum_{i=1}^{n\minus j}f(i)
		\nonumber \\ && + \sum_{j=0}^{n\minus 1}\frac{(m\plus j)!}{j!\ m}f(n\minus j)
	\\
	\sum_{j=0}^n\frac{(m\minus 1\plus j)!}{j!}S_1(m\plus j)f(n\plus 1\minus j) & = &
		-(m\minus 1)\sum_{j=0}^{n\plus 1}\frac{(m\minus 2\plus j)!}{j!}S_1(m\minus 1\plus j)f(n\plus 2\minus j)
	\nonumber \\ &&
		+\sum_{j=0}^{n\plus 1}\frac{(m\minus 1\plus j)!}{j!}S_1(m\minus 1\plus j)f(n\plus 2\minus j)
\end{eqnarray}
This is used for the next equations
\begin{eqnarray}
	\sum_{j=0}^{n\minus 1}\binom(m\plus j,j)\frac{S_1(m\plus j)}{n\minus j} & = &
	\binom(m\plus n,n)(2S_{1,1}(m\plus n)-2S_2(m\plus n)
		\nn
	+S_2(m)-S_1(m\plus n)S_1(m))
	\\
	\sum_{j=0}^{n\minus 1}\binom(m\plus j,j)S_1(m\plus j)S_1(n\minus j) & = &
		\binom(n\plus m\plus 1,n)(-(S_2(n\plus m\plus 1)-S_2(m\plus 1))
		\nonumber \\ &&
			+(S_1(n\plus m\plus 1)\minus S_1(m\plus 1))
				(S_1(n\plus m\plus 1)\minus\frac{1}{m\plus 1}))
\end{eqnarray}

\subsection{Sums with $\sign(j)$}
 
The next sums all contain a factor $(\minus 1)^j$ and hence they give
completely different results than the corresponding set of sums without
the $(\minus 1)^j$. The most important ones have been treated in the 
appendices \ref{ap:conjugations} and \ref{ap:nmini}.

\begin{eqnarray}
	\sum_{j=0}^n\binom(n,j)(\minus 1)^j\frac{1}{m\plus j} & = &
			\binom(n\plus m,n)^{-1}\frac{1}{m}
\\
	\sum_{j=0}^n\binom(n,j)(\minus 1)^jS_1(m\plus j) & = &
			-\binom(n\plus m,n)^{-1}\frac{1}{n}
\\
	\sum_{j=0}^n\binom(n,j)(\minus 1)^j\frac{1}{(m\plus j)^2} & = &
			\binom(n\plus m,n)^{-1}\frac{1}{m}
			(S_1(m\plus n)-S_1(m\minus 1))
\\
	\sum_{j=0}^n\binom(n,j)(\minus 1)^j\frac{S_1(m\plus j)}{m\plus j} & = &
			\binom(n\plus m,n)^{-1}\frac{1}{m}
			(S_1(m\plus n)-S_1(n))
\\
	\sum_{j=0}^n\binom(n,j)(\minus 1)^jS_2(m\plus j) & = &
			-\binom(n\plus m,n)^{-1}\frac{1}{n}
			(S_1(n\plus m)-S_1(m))
\\
	\sum_{j=0}^n\binom(n,j)(\minus 1)^jS_{1,1}(m\plus j) & = &
			-\binom(n\plus m,n)^{-1}\frac{1}{n}
			(S_1(n\plus m)-S_1(n\minus 1))
\end{eqnarray}
\begin{eqnarray}
	\sum_{j=0}^n(\minus 1)^j\binom(n,j)\frac{S_1(n\minus j)}{m\plus j} & = &
		-A_1(n,m)(\ (n\plus m)\binom(n\plus m\minus 1,n)\ )^{-1}
\end{eqnarray}

There is also a number of sums with more than one binomial coefficient. 
First a few general ones.
\begin{eqnarray}
	\sum_{j=0}^n(\minus 1)^j\binom(n,j)\binom(m\plus j,j) & = &
			(\minus 1)^n\binom(m,n)\ \ m \ge n
\\
	\sum_{j=0}^n(\minus 1)^j\binom(n,j)\binom(m\plus j,j) & = & 0
		\ \ m < n
\\
	\sum_{j=0}^n(\minus 1)^j\binom(n,j)\binom(m\plus j,j)\ j & = &
			(\minus 1)^n\binom(m\plus 1,n)\ n\ \ m \ge n\minus 1
	\nonumber \\ & = & 0 \ \ \ \ \ \ \ \ \ \ \ \ m < n\minus 1
\\
	\sum_{j=0}^n\binom(n,j)\binom(n\plus j,m\plus j)(\minus 1)^j & = & 0\ \ \ 0 < m < n
\\
	\sum_{j=0}^n\binom(n,j)\binom(n\plus m\plus j,m\plus j)(\minus 1)^j & = & (\minus 1)^n\ \ \ m \ge 0
\\
	\sum_{j=0}^n\frac{1}{j!(n\plus 1\minus j)!}\frac{(2n\plus 2\minus j)!}{(n\plus 1\minus j)!}(\minus 1)^j & = &
			1+(\minus 1)^n
\\
	\sum_{j=0}^{n\minus 1}\frac{1}{j!(n\minus 1\minus j)!}\frac{(n\minus 1\plus j)!}{j!(j\plus 2)}(\minus 1)^j & = &
			\frac{1}{2}\delta(n\minus 1)-\frac{1}{6}\delta(n\minus 2)
\end{eqnarray}
And now the sums with two binomial coefficients and an argument j in the 
extra piece (see also appendix \ref{ap:miscel}):
\begin{eqnarray}
	\sum_{j=1}^n\binom(n,j)\binom(n\plus j,j)(\minus 1)^jS_{1,1}(j) & = &
				 \sign(n)(4S_{1,1}(n)-2S_2(n))
\\
	\sum_{j=1}^n\binom(n,j)\binom(n\plus j,j)(\minus 1)^j\frac{1}{j}S_1(j) & = &
				 2S_{\minus 2}(n)
\\
	\sum_{j=1}^n\binom(n,j)\binom(n\plus j,j)(\minus 1)^jS_{1,2}(j) & = &
				\sign(n)(2S_{\minus 3}(n)-4S_{1,\minus 2}(n))
\\
	\sum_{j=1}^n\binom(n,j)\binom(n\plus j,j)(\minus 1)^jS_{2,1}(j) & = &
				\sign(n)(2S_3(n))
\\
	\sum_{j=1}^n\binom(n,j)\binom(n\plus j,j)(\minus 1)^jS_{1,1,1}(j) & = &
				\sign(n)(2S_3(n)-4S_{1,2}(n)-4S_{2,1}(n)+8S_{1,1,1}(n))
\\
	\sum_{j=1}^n\binom(n,j)\binom(n\plus j,j)(\minus 1)^j\frac{1}{j}S_2(j) & = &
				 -2S_3(n)
\\
	\sum_{j=1}^n\binom(n,j)\binom(n\plus j,j)(\minus 1)^j\frac{1}{j}S_{1,1}(j) & = &
				 4S_{\minus 2,1}(n)-2S_{\minus 3}(n)
\\
	\sum_{j=1}^n\binom(n,j)\binom(n\plus j,j)(\minus 1)^j\frac{1}{j^2}S_1(j) & = &
				 4S_{1,\minus 2}(n)-2S_{\minus 3}(n)
\end{eqnarray}
Similarly there are formulae with $j+1$ (see also appendix \ref{ap:miscel}).
\begin{eqnarray}
	\sum_{j=0}^n\binom(n,j)\binom(n\plus j,j\plus 1)\sign(j)S_1(j\plus 1) & = &
				\sign(n)\frac{1}{n\plus 1}
\\
	\sum_{j=0}^n\binom(n,j)\binom(n\plus j,j\plus 1)\sign(j)
				\frac{S_1(j\plus 1)}{j\plus 1} & = &
				-\sign(n)\frac{1}{n(n\plus 1)^2}
\\
	\sum_{j=0}^n\binom(n,j)\binom(n\plus j,j\plus 1)\sign(j)S_{1,1}(j\plus 1) & = &
				\frac{\sign(n)}{n\plus 1}(
					S_1(n\plus 1)+S_1(n\minus 1))
\\
	\sum_{j=0}^n\binom(n,j)\binom(n\plus j,j\plus 1)\sign(j)S_2(j\plus 1) & = &
				-\frac{1}{n(n\plus 1)^2}
\\
	\sum_{j=0}^n\binom(n,j)\binom(n\plus j,j\plus 1)\sign(j)
			\frac{S_1(j\plus 1)}{(j\plus 1)^2} & = &
		-\frac{1}{n\plus 1}(S_{\minus 2}(n\plus 1)+S_{\minus 2}(n\minus 1)
			+\frac{\sign(n)}{n(n\plus 1)})
\\
	\sum_{j=0}^n\binom(n,j)\binom(n\plus j,j\plus 1)\sign(j)
				\frac{S_{1,1}(j\plus 1)}{j\plus 1} & = &
			\frac{\sign(n)}{n(n\plus 1)^2}(1-S_1(n\plus 1)
					-S_1(n\minus 1))
\\
	\sum_{j=0}^n\binom(n,j)\binom(n\plus j,j\plus 1)\sign(j)
			\frac{S_2(j\plus 1)}{j\plus 1} & = &
				\frac{1}{n^2(n\plus 1)^2}+\frac{1}{(n\plus 1)^3}
\\
	\sum_{j=0}^n\binom(n,j)\binom(n\plus j,j\plus 1)\sign(j)S_3(j\plus 1) & = &
			\frac{1}{n(n\plus 1)^2}(1-S_1(n\plus 1)
					-S_1(n\minus 1))
\end{eqnarray}
And the formulae with $j+2$ (see also appendix \ref{ap:miscel}).
\begin{eqnarray}
	\sum_{j=0}^n\binom(n,j)\binom(n\plus j,j\plus 2)\sign(j) S_1(j\plus 2) & = &
				-\sign(n)\frac{1}{(n\plus 1)(n\plus 2)}
\\
	\sum_{j=0}^n\binom(n,j)\binom(n\plus j,j\plus 2)\sign(j) S_2(j\plus 2) & = &
				-\frac{2(n^2\plus n\plus 1)}{(n\minus 1)n(n\plus 1)^2(n\plus 2)^2}
\\
	\sum_{j=0}^n\binom(n,j)\binom(n\plus j,j\plus 2)\sign(j) S_3(j\plus 2) & = &
			\frac{1}{(n\plus 1)(n\plus 2)}(
				\frac{11/6}{n\minus 1}-\frac{5/2}{n}
				+\frac{5/2}{n\plus 1}-\frac{11/6}{n\plus 2} + \sign(n)\times
		\nonumber \\ &&
				\times(S_1(n\plus 2)+S_1(n\minus 2))
          (S_{\minus 1}(n\plus 2)-S_{\minus 1}(n\minus 2))\ )
\\
	\sum_{j=0}^n\binom(n,j)\binom(n\plus j,j\plus 2)\sign(j)
			\frac{S_2(j\plus 2)}{j\plus 2} & = &
			\frac{1}{(n\plus 1)(n\plus 2)}(S_2(n\plus 2)-S_2(n\minus 2)
		\nonumber \\ &&
				-\frac{5}{6}\frac{1}{n\minus 1}+\frac{1}{2}\frac{1}{n}
				-\frac{1}{2}\frac{1}{n\plus 1}+\frac{5}{6}\frac{1}{n\plus 2})
\end{eqnarray}
The formulae with $j+n$:
\begin{eqnarray}
	\sum_{j=0}^n\binom(n,j)\binom(n\plus j,j)\sign(j)\frac{1}{n\plus j} & = &
				 0
\\
	\sum_{j=0}^n\binom(n,j)\binom(n\plus j,j)\sign(j)S_1(n\plus j) & = &
				 2\sign(n)S_1(n)
\\
	\sum_{j=0}^n\binom(n,j)\binom(n\plus j,j)\sign(j)\frac{1}{(n\plus j)^2} & = &
				-\sign(n)\frac{(n\minus 1)!(n\minus 1)!}{(2n)!}
\\
	\sum_{j=0}^n\binom(n,j)\binom(n\plus j,j)\sign(j)(
			S_{1,1}(n\plus j)-S_2(n\plus j)) & = &
				\sign(n)(4S_{1,1}(n)-3S_2(n))
\end{eqnarray}
\begin{eqnarray}
	\sum_{j=1}^n\binom(n,j)\binom(n\plus j,j)\sign(j)\frac{1}{j}S_1(n\plus j) & = &
				3S_2(n)+2S_{\minus 2}(n)-4S_{1,1}(n)
\\
	\sum_{j=0}^n\binom(n,j)\binom(n\plus j,j)\sign(j)\frac{1}{j\plus 1}S_1(n\plus j) & = &
				\sign(n)\frac{1}{n(n\plus 1)}
\\
	\sum_{j=0}^n\binom(n,j)\binom(n\plus j,1\plus j)\sign(j)S_1(n\plus j) & = &
			\sign(n)\frac{1}{n\plus 1}
\\
	\sum_{j=0}^n\binom(n,j)\binom(n\plus j,1\plus j)\sign(j)S_2(n\plus j) & = &
			\sign(n)\frac{n!\ (n\minus 1)!}{(2n)!}\frac{1}{n\plus 1}
\\
	\sum_{j=0}^n\binom(n,j)\binom(n\plus j,2\plus j)\sign(j)S_2(n\plus j) & = &
			\sign(n)\frac{n!\ (n\minus 1)!}{(2n)!}(
				\frac{1}{3}\frac{1}{n\plus 2}-\frac{1}{n\plus 1}-\frac{1}{3}\frac{1}{n\minus 1})
\\
	\sum_{j=0}^n\binom(n,j)\binom(n\plus j,3\plus j)\sign(j)S_2(n\plus j) & = &
			\sign(n)\frac{n!\ (n\minus 1)!}{(2n)!}(
				\frac{1}{10}\frac{1}{n\plus 3}-\frac{2}{3}\frac{1}{n\plus 2}
	\nonumber \\ && \ \ \ \ \ \ +\frac{1}{n\plus 1}
			+\frac{1}{6}\frac{1}{n\minus 1}+\frac{2}{5}\frac{1}{n\minus 2})
\\
	\sum_{j=0}^n\binom(n,j)\binom(n\plus j,j)\sign(j)
					\frac{1}{(j\plus 1)^2}S_1(n\plus j) & = &
			\frac{1}{n(n\plus 1)}S_1(n\minus 1)-\sign(n)\frac{1}{n^2(n\plus 1)^2}
\\
	\sum_{j=0}^n\binom(n,j)\binom(n\plus j,j)\sign(j)\frac{1}{j\plus 2}S_1(n\plus j) & = &
				\sign(n)(\frac{1}{6}\frac{1}{n\minus 1}
					+\frac{1}{2}\frac{1}{n}-\frac{1}{2}\frac{1}{n\plus 1}
					-\frac{1}{6}\frac{1}{n\plus 2})
\\
	\sum_{j=0}^n\binom(n,j)\binom(n\plus j,2\plus j)\sign(j)\frac{1}{j\plus 2}S_1(n\plus j) & = &
			\frac{n!}{(n\plus 2)!}(S_1(n\minus 2)
				+2\sign(n)(n^2\plus n\plus 1)\frac{(n\minus 2)!}{(n\plus 2)!})
\\
	\sum_{j=0}^n\binom(n,j)\binom(n\plus j,3\plus j)\sign(j)\frac{1}{j\plus 3}S_1(n\plus j) & = &
			\frac{2\ n!}{(n\plus 3)!}(S_1(n\minus 3)
				-(S_{\minus 1}(n\plus 3)-S_{\minus 1}(n\minus 3))\ )
\\
	\sum_{j=0}^n\binom(n,j)\binom(n\plus j,1\plus j)\sign(j)
					\frac{1}{(j\plus 1)^2}S_1(n\plus j) & = &
			\frac{1}{n\plus 1}(
				2S_{1,1}(n\minus 1)-2S_2(n\minus 1)
					-S_{\minus 2}(n\plus 1)
		\nonumber \\ &&
			-S_{\minus 2}(n\minus 1)
			+S_1(n\plus 1)S_1(n\minus 1)
			-\frac{\sign(n)}{n(n\plus 1)})\ )
\\
	\sum_{j=0}^n\binom(n,j)\binom(n\plus j,2\plus j)\sign(j)
					\frac{1}{(j\plus 2)^2}S_1(n\plus j) & = &
			\frac{n!}{(n\plus 2)!}(
				2S_{1,1}(n\minus 2)-2S_2(n\minus 2)-S_1(n\minus 2)
		\nonumber \\ &&
				-S_{\minus 2}(n\plus 2)-S_{\minus 2}(n\minus 2)
				+S_1(n\plus 2)S_1(n\minus 2)
		\nonumber \\ &&
			+\sign(n)(n^2\plus n\plus 3)\frac{(n\minus 2)!}{(n\plus 2)!}\ )
\end{eqnarray}
And some mixed formulae.
\begin{eqnarray}
	\sum_{j=0}^n\binom(n,j)\binom(n\plus j,j)(\minus 1)^j\frac{1}{m\plus j} & = &
				0\ \ \ \ \ \ \ \ \ \ \ \ \ \ \ \ \ \ \ \ (m \le n)
		\nonumber \\ & = &
				(\minus 1)^n\frac{(m\minus 1)!(m\minus 1)!}{(n\plus m)!
				(m\minus n\minus 1)!}\ \ \ (m > n)
\\
	\sum_{j=0}^n\binom(n,j)\binom(n\plus j,j)(\minus 1)^j\frac{1}{(m\plus j)^2} & = &
				-(\minus 1)^m\frac{(m\minus 1)!(m\minus 1)!(n\minus m)!}{(n\plus m)!}\ \ \ (n \ge m)
\\
	\sum_{j=0}^n\binom(n,j)\binom(n\plus j,m\plus j)(\minus 1)^j S_2(m\plus j) & = &
				(\minus 1)^{n+m}\frac{(m\minus 1)!\ n!}{(n\plus m)!}
				(S_{\minus 1}(n\plus m)-S_{\minus 1}(n\minus m))
		\nonumber \\ && \ \ \ \ \ \ \ \ \ \ \ \ \ \ \ \ \ \ \ \ 
				\ \ \ (n \ge m)
\\
	\sum_{j=0}^n\binom(n,j)\binom(n\plus j,m\plus j)(\minus 1)^jS_1(n\plus j) & = &
			-(\minus 1)^{n\plus m}\frac{n!\ (m\minus 1)!}{(n\plus m)!}
				\ \ \ \ \ \ \ \ \ (n \ge m)
\end{eqnarray}

There are also sums with two factorials in the numerator.
\begin{eqnarray}
	\sum_{j=0}^nj!(m\minus j)!(\minus 1)^j & = & \frac{(m\plus 1)!}{m\plus 2}+(\minus 1)^n
				\frac{(n\plus 1)!\ (m\minus n)!}{m\plus 2}
\\
	\sum_{j=1}^nj!\ (m\minus j)!(\minus 1)^jS_1(j) & = &
			(\minus 1)^n\frac{(n\plus 1)!\ (m\minus n)!}{m\plus 2}
			(S_1(n\plus 1)-\frac{1}{m\plus 2})
	\nonumber \\ &&
			-\frac{(m\plus 1)!}{(m\plus 2)^2}
\\
	\sum_{j=0}^nj!\ (n\minus j)!(\minus 1)^j\frac{1}{j\plus 1} & = &
			-(n\plus 1)!\ (S_2(n\plus 1)+2S_{\minus 2}(n\plus 1))
\\
	\sum_{j=1}^nj!\ (n\minus j)!(\minus 1)^j\frac{1}{j^2} & = &
			n!\ (S_2(n)+2S_{\minus 2}(n))
\\
	\sum_{j=0}^m\sign(j)j!\frac{(n\plus m\minus j)!}{2\plus m\minus j} & = &
		\sign(m)(m\plus 2)!\ (n\minus 2)!\ (A_1(m\plus 3,n\minus 2)
						+S_1(m\plus 2))
	\nonumber \\ &&
		+\frac{(m\plus n\plus 1)!}{(m\plus 3)^2}
		+\sign(m)(m\plus 1)!\ (n\minus 1)!
\\
	\sum_{j=0}^n\sign(j)(n\plus m\minus j)!\ j!\ A_1(j,n\minus j) & = &
		\frac{(n\plus m\plus 1)!}{n\plus m\plus 2}(S_1(n\plus m\plus 1)
			-S_1(m\plus 1))
	\nonumber \\ &&
			-\sign(n)\frac{(n\plus 1)!\ m!}{n\plus m\plus 2}S_1(n)
\end{eqnarray}

\subsection{Partial sums}

For derivations the use of partial sums (sums of only part of the range of 
the binomial coefficients) are very useful. Unfortunately these are hard to 
obtain and hence only a limited number of them can be presented here.
\begin{eqnarray}
	\sum_{j=0}^m(\minus 1)^j\binom(n,j) & = & (\minus 1)^m\binom(n\minus 1,m)
\\
\label{eq:defA}
	\sum_{j=1}^m(\minus 1)^j\binom(n,j)\frac{1}{j^k} & = &
		A_k(m,n\minus m)
\\
	\sum_{j=0}^m(\minus 1)^j\binom(n,j)S_1(j) & = &
		(\minus 1)^m\binom(n\minus 1,m)(S_1(m)+\frac{1}{n})-\frac{1}{n}
\\
	\sum_{j=0}^m(\minus 1)^j\binom(n,j)S_1(n\minus j) & = &
		(\minus 1)^m\binom(n\minus 1,m)(S_1(n\minus m\minus 1)+\frac{1}{n})
\end{eqnarray}

%

\end{document}